\documentclass[twocolumn,twocolappendix]{aastex701}

\usepackage{natbib}

\usepackage{adjustbox}
\usepackage{comment}
\usepackage{graphicx}	% Including figure files
\usepackage{CJK}

\usepackage{amsmath,bm}	% Advanced maths commands
\usepackage{wasysym}    % Extra astronomy symbols
\usepackage{xspace}
\usepackage[version=4]{mhchem}
\usepackage{soul}
\usepackage{txfonts,textcomp}

\newcommand\Msun{\text{M}_{\astrosun}} % requires the wasysym package
\newcommand\Zsun{\text{Z}_{\astrosun}} % requires the wasysym package

\newcommand\arepo{\textsc{arepo}\xspace}

\begin{document}
\begin{CJK*}{UTF8}{gbsn}

%\title{Linking Stellar Masses and their Parental Cluster in High Resolution Simulations of Galaxies}
\title{Individual Star Sampling in Star Formation Simulations: A Semi-Deterministic Model}

\author[orcid=0000-0002-7478-6427,sname='Deng']{Yunwei Deng (邓云未)}
\affiliation{Department of Astronomy, Tsinghua University, Haidian DS 100084, Beijing, China}
\email{dengyw22@mails.tsinghua.edu.cn}  

\author[orcid=0000-0002-1253-2763, sname='Li']{Hui Li (李辉)} 
\affiliation{Department of Astronomy, Tsinghua University, Haidian DS 100084, Beijing, China}
\email[show]{hliastro@tsinghua.edu.cn}

\author[orcid=0000-0001-7395-1198,sname='Yan']{Zhiqiang Yan (闫智强)}
\affiliation{School of Astronomy and Space Science, Nanjing University, Nanjing, 210093, China}
\affiliation{Key Laboratory of Modern Astronomy and Astrophysics, Nanjing University, Ministry of Education, Nanjing 210093, China}
\email{yan@nju.edu.cn}

\author[orcid=0009-0003-9099-5805,sname='Kong']{Chuizheng Kong (孔垂政)}
\affiliation{Department of Astronomy, Tsinghua University, Haidian DS 100084, Beijing, China}
\email{kcz24@mails.tsinghua.edu.cn}

\author[orcid=0000-0002-7299-2876,sname='Zhang']{Zhi-Yu Zhang (张智昱)}
\affiliation{School of Astronomy and Space Science, Nanjing University, Nanjing, 210093, China}
\affiliation{Key Laboratory of Modern Astronomy and Astrophysics, Nanjing University, Ministry of Education, Nanjing 210093, China}
\email{zzhang@nju.edu.cn}
% \author[sname=' Grudi\'c']{Michael Y. Grudi\'c}
% \affiliation{South African Astronomical Observatory}
% \affiliation{University of Cape Town, Department of Astronomy}
% \email{fakeemail3@google.com}

%% Use the \collaboration command to identify collaborations. This command
%% takes an optional argument that is either a number or the word "all"
%% which tells the compiler how many of the authors above the command to
%% show. For example "\collaboration[all]{(DELVE Collaboration)}" wil include
%% all the authors above this command.
%%
%% Mark off the abstract in the ``abstract'' environment. 
\begin{abstract}
In modern simulations that include star formation, it is common to use a universal and invariant initial mass function (IMF) to represent star populations or sample individual stars. However, stellar masses are determined by local and environmental processes that operate over a wide dynamical range and remain unresolved in simulations. We introduce a semi-deterministic (SDT) scheme for sampling individual stars from star-forming gas in numerical simulations. We represent unresolved molecular cores and protostellar disks with reservoir particles (RsvPs) and employ an on-the-fly friends-of-friends algorithm to identify star clusters. The instantaneous IMF for newly formed stars is then derived from the current cluster mass. We test the performance of this method in simulations of isolated molecular clouds and a major merger between two dwarf galaxies. Compared to existing IMF sampling methods, our SDT scheme naturally reproduces the observed $m_{\star,\text{max}}$-$M_\text{ecl}$ relation and yields numbers of massive stars consistent with optimal sampling theory. It also exhibits the smallest run-to-run variation among simulations with different random seeds. The regulated star formation results in a small ($\sim0.15$\,Myr) but coherent time delay in the emergence of massive stars, reduces the large scatter arising from Poisson noise, and produces initial mass segregation within the clusters. On galactic scales, the SDT method predicts a steeper high-mass IMF slope at low star formation rates (SFRs), with the slope negatively correlated with the SFR. As the specific abundance of massive stars declines, our model naturally explains the systematically low H$\alpha$-based SFRs, where the H$\alpha$ indicator for Milky Way-like galaxies underestimate the intrinsic SFR owing to the reduced population of massive stars.

\end{abstract}

%% Keywords should appear after the \end{abstract} command. 
%% The AAS Journals now uses Unified Astronomy Thesaurus (UAT) concepts:
%% https://astrothesaurus.org
%% You will be asked to selected these concepts during the submission process
%% but this old "keyword" functionality is maintained in case authors want
%% to include these concepts in their preprints.
%%
%% You can use the \uat command to link your UAT concepts back its source.

%% From the front matter, we move on to the body of the paper.
%% Sections are demarcated by \section and \subsection, respectively.
%% Observe the use of the LaTeX \label
%% command after the \subsection to give a symbolic KEY to the
%% subsection for cross-referencing in a \ref command.
%% You can use LaTeX's \ref and \label commands to keep track of
%% cross-references to sections, equations, tables, and figures.
%% That way, if you change the order of any elements, LaTeX will
%% automatically renumber them.

\section{Introduction}
\label{sec:intro}
Stars are the fundamental building blocks of galaxies. Through their formation, evolution, and death, they emit radiation, release heavy elements, and inject energy and momentum, thereby driving the baryonic cycle of the Universe. Among all stars, massive stars play a particularly important role. Although they constitute only a small fraction of the stellar population by number, they dominate the production of ionizing radiation, stellar winds, supernovae (SNe), and heavy elements \citep{2013ApJ...770...25A,2019ARA&A..57..227K}. Consequently, the number, masses, and formation times of massive stars strongly influence the life cycle of star-forming clouds \citep{2024ARA&A..62..369S}, as well as the star formation histories, chemical enrichment, and dynamical evolution of galaxies \citep{2018MNRAS.480.1666S,2020MNRAS.492....8A,2024A&A...681A..28A}. Therefore, determining how stellar masses are assigned during star formation is a fundamental problem in simulations of molecular cloud (MC) evolution and galaxy formation.

The influence of stellar mass sampling becomes especially significant in low-mass environments, such as dwarf galaxies and individual molecular clouds, which serve as the progenitors of more-massive systems in a hierarchical $\Lambda$ cold dark matter ($\Lambda$CDM) framework. In these environments, only a small number of massive stars are expected to form, making stellar feedback highly sensitive to the details of the stellar mass assignment \citep[e.g.][]{2011ApJ...741L..26F,2021MNRAS.502.5417S,2024arXiv241117862J:Jeon}. For example, \citet{2023MNRAS.526.1713S:Steyrleithner} showed that modifying the truncation of the initial mass function (IMF) of massive stars in dwarf galaxies changes their SN rate, outflow properties, and star formation history. Since low-mass galaxies are dominated by low-mass star clusters and exhibit intrinsically low star formation rates (SFRs), they are particularly sensitive to the IMF sampling prescriptions in numerical simulations.

Recently, galaxy formation simulations with solar-mass resolution have recently begun resolving individual stars with discrete stellar masses \citep[e.g.,][]{2017MNRAS.471.2151H,2020ApJ...891....2L,2021MNRAS.502.5417S,2024A&A...691A.231D:Deng,2025ApJ...984..142G,2026MNRAS.545f2158J:Jeon,2025ApJ...978..129A:Andersson,2026arXiv260400100S:Smith,2026arXiv260610000C:Chung}. This transition from IMF-averaged stellar populations to explicit star-by-star modeling has elevated stellar mass assignment to a central component of numerical galaxy formation models. However, even at solar-mass resolution, simulations cannot yet predict stellar masses from first principles. The physical processes that regulate stellar masses involve a wide range of scales, including cloud fragmentation, protostellar accretion, stellar dynamics, and feedback, many of which remain unresolved in current simulations.

In the absence of a fully \textit{ab initio} model, most simulations assume that stellar masses follow a universal stellar initial mass function \citep[IMF;][]{1955ApJ...121..161S}. The standard approach is to interpret the IMF as a probability density function and draw stellar masses independently from it. This procedure, commonly referred to as ``random'' (RND) or ``stochastic'' IMF sampling, is computationally simple and naturally reproduces the IMF when averaged over a sufficiently large stellar population.

\begin{figure}
\includegraphics[width=\columnwidth]{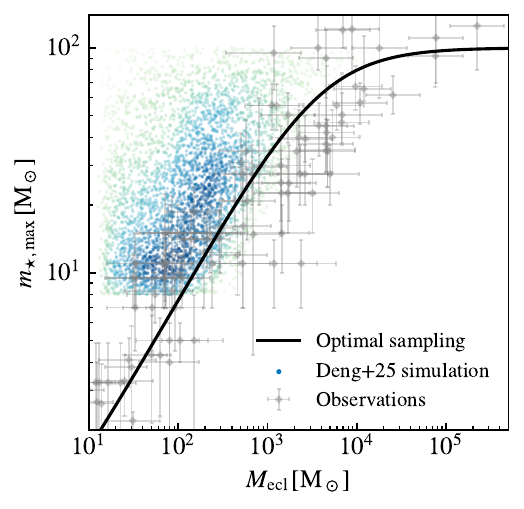}
\caption{Most-massive stellar mass to embedded cluster mass ($m_{\star,\text{max}}$--$M_\text{ecl}$) relation. The black curve is the theoretical relation predicted by the optimal sampling following the deviations in \cite{2023A&A...670A.151Y:Yan}. Blue dots are the results from the simulation of \cite{2025A&A...704A.240D:Deng} using the \cite{2024A&A...691A.231D:Deng} version RIGEL model, color-coded by the normalized density in log-space estimated by a Gaussian kernel. In this simulation, $m_{\star,\text{max}}$ is allowed to exceed $M_\text{ecl}$, as the sampled massive star mass is not equal to the mass of star particles. Gray crosses are the observed young clusters (age $<5$\,Myr) from \cite{2011ApJ...727...64K}, \cite{2013MNRAS.434...84W:Weidner}, and \cite{2017ApJ...834...94S:Stephens}.}
\label{fig:mms}
\end{figure}

While technically straightforward, purely stochastic sampling faces growing observational challenges. One of the strongest constraints comes from the relation between the total stellar mass of an embedded cluster ($M_\mathrm{ecl}$) and the mass of its most-massive star ($m_{\star,\mathrm{max}}$) \citep{2003ASPC..287...65L,2006MNRAS.365.1333W:Weidner,2010MNRAS.401..275W:Weidner,2023A&A...670A.151Y:Yan}. As illustrated in Fig.~\ref{fig:mms}, low-mass clusters can only host less-massive stars. For example, a cluster with $M_\mathrm{ecl}\sim100\,\Msun$ rarely contains stars above $20\,\Msun$, whereas clusters with $M_\mathrm{ecl}\sim10^4\,\Msun$ can form stars approaching $100\,\Msun$. Statistically, observations of young clusters and associations show substantially smaller scatter in both the fraction of massive stars and the mass of the most-massive star than expected from repeated independent draws from a Salpeter-like IMF \citep{2002Sci...295...82K,2023A&A...670A.151Y:Yan,2025MNRAS.538.2989C}. Using an updated observational sample, \cite{2023A&A...670A.151Y:Yan} showed that the observed $m_{\star,\mathrm{max}}$--$M_\mathrm{ecl}$ relation is inconsistent with pure random sampling at a significance level exceeding $4.5\sigma$.

Likewise, high-resolution magnetohydrodynamic simulations of clustered star formation produce upper-mass distributions that are significantly narrower than predicted by stochastic sampling from an invariant IMF \citep{2023OJAp....6E..48G}. These results suggest that local gas availability, cloud structure, gravitational collapse, and stellar feedback introduce environmental constraints that suppress the extreme fluctuations expected from independent sampling. The resulting correlation between stellar masses and their parent clusters also affects the integrated observational properties of galaxies, including the H$\alpha$-to-far-UV (FUV) luminosity ratio \citep{2007ApJ...671.1550P:Pflamm-Altenburg,2009MNRAS.395..394P:Pflamm-Altenburg,2009ApJ...706..599L,2012ApJ...744...44W:Weisz}, the mass-to-light ratio \citep{2011MNRAS.415.1647G:Gunawardhana,2025A&A...702A.229R}, the photometric and spectroscopic properties \citep{2021A&A...655A..19Y,2024A&A...689A.221H:Haslbauer}, the mass--metallicity relation \citep[][]{2007MNRAS.375..673K:Koppen}, and $\alpha$-element enhancement relative to the Milky Way \citep[][]{2020A&A...637A..68Y:Yan,2021NatAs...5.1247M}.

Despite these indications, most galaxy formation simulations still adopt purely stochastic IMF sampling. This can lead to substantial discrepancies in low-mass clusters, where the highest stellar masses are dominated by small number statistics. For example, the blue points in Fig.~\ref{fig:mms} show clusters from the simulation of \citet{2025A&A...704A.240D:Deng}. In these simulations, the most-massive stars are systematically overproduced in low-mass clusters relative to the observed relation, with the discrepancy becoming increasingly severe toward smaller cluster masses. Because low-mass clusters dominate star formation in dwarf galaxies, this issue is expected to have its largest impact precisely in the regime where stellar feedback is most sensitive to the presence of massive stars.

In this work, we propose a {\it semi-deterministic} (SDT) split-sampling method applied in hydrodynamical galaxy simulations to link stellar masses and their parental clusters. Its core logic is to treat the sampling process differentially: For the critical few that dominate system volatility, i.e., the most-massive stars, we abandon simple random drawing. Instead, we employ deterministic assignment based on observational constraints. For the vast majority of low- and intermediate-mass stars that constitute the IMF body, we retain random sampling to save computational costs of time and memory.

Our applied deterministic sampling method for the most-massive stars is the ``optimal sampling'' \citep{2013pss5.book..115K:Kroupa}, which assumes that the formation of all of the stars in a star cluster is strongly correlated. By definition, optimal sampling distributes the physical reservoir $M_\text{ecl}$ according to a predefined IMF, such that the entire mass range is covered. Therefore, it is a deterministic method to populate the IMF. Although the physical necessity for the deterministic IMF sampling remains debated (e.g. \citealt{2023MNRAS.525.6182S}; however, see \citealt{2026RAA....26e5012G:Gjergo} for a recent theoretical discussion), optimal sampling does provide a good interpretation of the observed $m_{\star,\text{max}}$--$M_\text{ecl}$ relation, as the black curve in Fig.~\ref{fig:mms} shows.

To connect the most-massive stellar mass with the growing total stellar mass formed in a star cluster, we use efficient real-time friends-of-friends (FoF) clustering algorithms to enable each newly formed star to access their environmental information and adopt optimal sampling theory as a tool to determine the upper mass limit for IMF sampling. Therefore, our SDT method can tightly correlate the highest stellar mass with their parent cluster mass in high-resolution simulations of galaxies, in agreement with observations.

The remainder of this paper is organized as follows. In Section~\ref{sec:models}, we briefly introduce the phenomenological models for the $m_{\star,\text{max}}$--$M_\text{ecl}$ relation. In Section~\ref{sec:method}, we elaborate on the details of the SDT IMF sampling method. Using our SDT method as a reference, we evaluate the performance and compare the results with those obtained using RND sampling and the NGB methods in seven individual molecular clouds (Section~\ref{sec:gmcs}) and a merger system of two dwarf galaxies (Section~\ref{sec:gal}). Finally, we discuss our results in Section~\ref{sec:diss} and summarize in Section~\ref{sec:sum}.

\section{Phenomenological models for the $m_{\star,\text{max}}$--$M_\text{ecl}$ relation}
\label{sec:models}
\label{sec:opt-sample}

The physical origin of the $m_{\star,\text{max}}$--$M_\text{ecl}$ relation remains uncertain, and a definitive investigation would require spatial resolution far beyond the capabilities of our current simulations, as there is roughly a 9 order-of-magnitude difference between the physical scale of single-star formation and that of an entire galaxy. Addressing this problem is, therefore, not the aim of the present work. Instead, we provide an overview of commonly used phenomenological descriptions that can be applied to reproduce the observed $m_{\star,\text{max}}$--$M_\text{ecl}$ relation in our simulations.

The simplest model is an empirical fit to observational data. For example, \cite{2000ApJ...539..342E} and \cite{2003ASPC..287...65L} give single power-law relations of $m_{\star,\text{max}}\propto M_\text{ecl}^{\beta}$ with $\beta = 1/1.35$ and $0.45$, respectively. A piecewise power-law form can provide a more accurate description of the observed data \citep[e.g.,][]{2006MNRAS.365.1333W:Weidner}.

A widely used theoretical framework to interpret the observed $m_{\star,\text{max}}$--$M_\text{ecl}$ relation is optimal sampling \citep{2013pss5.book..115K:Kroupa}. Its essential assumptions are that (i) stars form in groups rather than in isolation \citep{2010MNRAS.404.1564P,2017ApJ...834...94S:Stephens}, so stellar masses should be assigned in connection with their embedded cluster, and (ii) star formation is strongly self-regulated \citep{2023OJAp....6E..48G,2023A&A...670A.151Y:Yan}, implying that the IMF can be treated as a deterministic mapping from a finite mass reservoir to a discrete set of stellar masses.

To outline how optimal sampling distributes stellar masses, we write the IMF as the piecewise function
\begin{equation}\label{eq:xi}
\xi(m) =
\begin{cases} 
0, & m<m_{\rm low}, \\
k_1 m^{-\alpha_1}, & m_{\rm low} \leqslant m<m_{\rm turn}, \\ 
k_2 m^{-\alpha_2}, & m_{\rm turn} \leqslant m<m_{\rm up}, \\
0, & m_{\rm up} \leqslant m.
\end{cases}
\end{equation}
\cite{2001MNRAS.322..231K} gave the parameters for the canonical IMF with $m_{\rm low}=0.08\,\Msun$, $m_{\rm turn}=0.5\,\Msun$, $m_{\rm up}=100\,\Msun$, $\alpha_1=1.3$, $\alpha_2=2.3$, and $k_1=2k_2$. The normalization of $\xi(m)$ is the total number of stars $N$ in a star cluster. We note that more general formalisms allow the slopes to depend on metallicity and gas clump density \citep[e.g.,][]{2021A&A...655A..19Y,2026RAA....26b5003G}. However, here we adopt fixed slopes to isolate the effects of IMF sampling.

Because optimal sampling assumes a stellar mass distribution without Poisson scatter, an explicit upper mass limit arises from the normalization condition. For a cluster with total stellar mass $M_{\rm ecl}$, the upper mass limit $m_\text{max}$ is generally lower than the assumed $m_{\rm up}$ in Eq.~\ref{eq:xi}, and is computed from
\begin{equation}\label{eq:mmax}
    1 = \int_{m_\text{max}}^{m_\text{up}}\xi(m){\rm d}m.
\end{equation}
Here, $m_\text{max}$ is determined simultaneously with the normalization parameters $k_1$ and $k_2$, subject to mass conservation and continuity of the IMF at $m_{\rm turn}$.
\begin{equation}\label{eq:norm}
    \begin{cases}
    M_\text{ecl} = &\int_{m_\text{low}}^{m_\text{max}}m\xi(m){\rm d}m\,,\\
    k_1m_\text{turn}^{-\alpha_1}=&k_2m_\text{turn}^{-\alpha_2}\,.
    \end{cases}
\end{equation}

Once $m_\text{max}$ and the normalization are fixed, the individual stellar masses can be obtained by partitioning the IMF into $N$ consecutive intervals, each containing exactly one star, and assigning each star the mass given by the first moment of $\xi(m)$ over its interval. We refer readers to \citet{2013pss5.book..115K:Kroupa} for the formal definition and to \citet{2015A&A...582A..93S:Schulz,2017A&A...607A.126Y:Yan,2023A&A...670A.151Y:Yan} for implementations in analytic models.

The phenomenological models correlate the masses of individual stars with the total stellar mass of their host cluster, $M_{\rm ecl}$. In hydrodynamical simulations, however, cluster members form gradually as the parent cloud collapses and fragments over several million years. Consequently, when the first stars appear, the final cluster mass is still unknown. At each time step, we can only estimate the mass of the stars that will form in the next step. In the following sections, we describe how this predicted stellar mass is calculated, and in Section~\ref{sec:sdt} we describe the additional procedures that ensure consistency with the $m_{\star,\text{max}}$--$M_\text{ecl}$ relation.

\section{Methods}
\label{sec:method}

In this section we describe how our SDT sampling framework is applied within the simulation. We first define the star-formation ``reservoir particles'' (RsvPs) used to represent unresolved collapsing cores (\hyperref[sec:rsvp-form]{Section~\ref*{sec:rsvp-form}}), then discuss how we enforce local mass conservation when forming stars at solar-mass resolution (\hyperref[sec:ngb]{Section~\ref*{sec:ngb}}). Next, we introduce our on-the-fly cluster identification procedure that provides an instantaneous estimate of $M_{\rm ecl}$ (\hyperref[sec:cl-find]{Section~\ref*{sec:cl-find}}), and finally present the SDT sampling algorithm that combines optimal and stochastic elements while remaining consistent with the $m_{\star,\text{max}}$--$M_{\rm ecl}$ relation (\hyperref[sec:sdt]{Section~\ref*{sec:sdt}}).
\subsection{Star formation reservoirs}
\label{sec:rsvp-form}

Although forecasting the final mass of the clusters in the simulation is challenging, it is at least straightforward to estimate the mass of the stars that will form in the next step. To do this, some simulations mark the gas particles about to form stars as ``star-forming particles.'' However, the star-forming gas can typically be very dense so that the unresolved or unmodeled physical processes, such as artificial fragmentation, optically thick cooling, and nonideal MHD, will be involved and result in an incorrect MHD solution. Thus, we adopt the concept ``reservoir particles (RsvPs)'' following the scheme outlined in \cite{2023MNRAS.522.3092L} and originally implemented by \cite{2019MNRAS.483.3363H} to treat these star-forming particles as a potential gas reservoir for future star formation, decoupled from the MHD solver. RsvPs represent dense star-forming cores crossing a series of thresholds.

A gas cell will turn into an RsvP if and only if it satisfies the following criteria:
\begin{enumerate}
    \item Jeans unstable threshold: The local Jeans length $L_\text{J}$ is resolved with less than $f_\text{J}$ gas cells, i.e. $L_\text{J}<f_\text{J}\Delta x_i$, where $\Delta x$ is the size of the gas cell $i$.
    \item Virial threshold: The gas cell is gravitationally unstable/self-gravitating at the resolution scale. The local virial parameter is evaluated by $[||\nabla {\bm v}_i||^2+(c_{s,i}/\Delta x)^2)]/8\pi G\rho_i<f_\alpha$, where $||\nabla {\bm v}_i||$ is the Frobenius norm of the gas velocity, $c_{s,i}$ is the sound speed, and $\rho_i$ is the gas density \citep{2013MNRAS.432.2647H,2019MNRAS.489.4233M}.
    \item Contracting flow threshold: The gas cell is a local sink where the gas flow is contracting, where $\nabla \cdot {\bm v_i}<0$.
    \item Density threshold: The gas cell is denser than a density threshold of $\rho_\text{th}$. Since the existence of the Jeans unstable threshold, the density threshold is maintained solely to avoid unphysical star formation in low-density regions. We take $\rho_\text{th}$ to be $0.1$ times the maximum density of the marginal Jeans resolved condition given by equation~(19) in \cite{2021MNRAS.506.2199G}.
    \item Temperature threshold: The temperature of the gas cell is lower than a threshold of $T_\text{th}$.
\end{enumerate}
Gas particles satisfying all five criteria will be converted to RsvPs immediately. RsvPs are decoupled from the hydrodynamics but are still involved with the gravity solver. The softening length of RsvPs ($\epsilon_{\rm RsvP}$) is assigned to be identical to that of gas particles at the time they are converted into RsvPs. The newly formed RsvPs will not form stars for a local dynamical time $t_\text{dyn}=f_\text{inert}t_\text{ff,0}$; we refer to these RsvPs as ``inert.'' Here, $t_\text{ff,0}=\sqrt{3\pi/(32G\rho)}$ is the freefall time evaluated with the density when the gas particle converts to RsvPs, and $f_\text{inert}$ is a scaling factor that defaults to unity. RsvPs older than their $t_\text{dyn}$ will begin to form stars.
Such RsvPs are referred to as ``active'' RsvPs. Conversely, RsvPs younger than their $t_\text{dyn}$ are designated as ``inert'' RsvPs.

\subsection{Local mass conservation in neighbor-based sampling}
\label{sec:ngb}
A long-standing difficulty for star formation in solar-mass resolution is local mass conservation: the direct IMF sampling methods \citep[e.g.,][]{2017MNRAS.466..407S:Sormani,2024A&A...691A.231D:Deng} usually allow massive stars (e.g. $20\,\Msun$) to be sampled from less-massive gas/sink/reservoir particles (e.g. $1\,\Msun$). Simulations focus on global or galaxy-scale properties and opt to ensure global mass conservation \citep[e.g.,][]{2015MNRAS.449..726F:Fujii,2024A&A...691A.231D:Deng}.
However, local mass conservation is fundamental to studying the internal structure and dynamics of clusters. 

\cite{2021PASJ...73.1036H} proposed a simple method to guaranty local mass conservation in the SIRIUS model. The candidate stellar mass is still stochastically sampled from a given IMF. After a target stellar mass is sampled, they examine whether there are sufficient gas reservoirs to actually form this star by summing up the total neighboring gas mass eligible to form stars within a maximum search radius $R_\text{search}$. If there are, they transfer the needed gas mass to the candidate star particle and convert it to a star. Otherwise, the candidate mass is discarded, and a new one is sampled.

The variants of this method are also used in the LYRA \citep{2021MNRAS.501.5597G} model and the latest iterations of the GRIFFIN model \citep{2023MNRAS.522.3092L,2025MNRAS.538.2129L:Lahen}. We call this family of sampling methods the ``neighbor-based'' or ``NGB'' method. The main differences among these models are the definitions of star formation reservoirs and the choice of the maximum searching radius $R_\text{search}$, which is inherently somewhat arbitrary. 

In our implementation of the NGB method, when mass transfer is necessary for forming a massive star, we begin by identifying neighboring RsvPs within the maximum search radius $R_\text{search}$ and creating a list of neighbors. This list is then sorted based on distance, and we merge the neighboring RsvPs into the target RsvP, proceeding from the closest to the farthest until we accumulate sufficient mass. The merged RsvPs inherit the center-of-mass coordinates and velocity of their progenitor RsvPs, while the new stars spawned from the RsvPs are spatially disturbed according to a Gaussian or uniform distribution with a standard deviation equal to the gravitational softening length $\epsilon_{\rm RsvP}$, with appropriate upper and lower limits (typically $0.1\epsilon_{\rm RsvP}$ and $10\epsilon_{\rm RsvP}$). 

This neighbor-based (NGB) sampling method, apart from ensuring mass conservation, does provide a way to link stellar mass with ambient environmental properties. \cite{2023MNRAS.522.3092L} compared the $m_{\star,\text{max}}$--$M_\text{ecl}$ relation obtained with this simple model with the observations compiled by \cite{2013MNRAS.434...84W:Weidner}. They found that the stacked relation in their simulations can produce the increasing trend between the highest stellar mass and the total mass of the cluster. However, individual low-mass clusters still exhibit strong volatility and can occasionally host a massive star with $m_{\star}\approx0.5M_\text{ecl}$. For the standard deviation of $m_{\star}$ the NGB results are barely distinguishable from those of the stochastic sample. 

Nonetheless, this method reproduces the essential behavior of the $m_{\star,\text{max}}$--$M_\text{ecl}$ relation while remaining straightforward to implement numerically. We, therefore, retain it as an option in our model and construct our SDT scheme on this basis by deterministically imposing the high-mass end using the instantaneous cluster mass.

A more stringent link between stellar masses and their parental clusters is difficult to establish in simulations, as we have no knowledge of the final mass of the cluster before the entire cluster is formed. In the subsequent subsections, we introduce our method based on on-the-fly cluster finding and address this problem.

\subsection{On-the-fly cluster finding}
\label{sec:cl-find}

Equipped with the $M_\text{ecl}$-dependent IMF sampling method (i.e., optimal sampling) and RsvPs, we now need to identify the ``clusters'' consisting of newly formed stars in the past and potential star-forming reservoirs in the near future.

To do this, we need to implement an on-the-fly clustering algorithm to link and group RsvPs and young star particles with physical connections, i.e. in the same cluster. For simplicity, we use the FoF algorithm. FoF uniquely defines groups that contain all particles separated by distance less than a given linking length $l_\text{link}$.

In principle, performing cluster finding at every time step provides the most up-to-date description of the evolving stellar clusters. However, FoF requires frequent communication among the computational tasks and nodes, making such updates expensive. Moreover, excessively frequent FoF updates are not always desirable, since too few newly formed RsvPs may have accumulated to assemble the newly required most-massive star. We, therefore, perform cluster finding at a fixed cadence of once every $\Delta t_\mathrm{FoF}$. $\Delta t_\mathrm{FoF}$ is a free parameter that determines how frequently the FoF groups are updated. If it is chosen too large, newly created RsvPs must wait an unnecessarily long time before being assigned to an FoF group, introducing an artificial temporal discretization into the star formation process. Conversely, if $\Delta t_\mathrm{FoF}$ is too small, too few newly formed RsvPs may have accumulated within an FoF group to assemble the newly required most-massive star predicted by the optimal sampling, thereby increasing the probability that the target stellar population cannot be fully realized. We, therefore, adopt $\Delta t_\mathrm{FoF}=0.1\,\mathrm{Myr}$. Since RsvPs are typically formed in bursts over approximately one local freefall time, this choice allows sufficient newly formed RsvPs to accumulate for assembling the optimally sampled stellar population while remaining short enough to avoid the artificial temporal discretization described above.

When conducting the FoF cluster finding, we only include these two types of particles:
\begin{enumerate}
    \item RsvPs that become active (i.e. older than $t_\text{dyn}$) before the next FoF point;
    \item the newly formed stars are younger than $\tau_\text{young}$. 
\end{enumerate}
$\tau_\text{young}$ is the age cutoff for young stars, and it is typically several Myr. An illustrative example demonstrating how young stars and RsvPs with different formation and activation times are grouped by the on-the-fly cluster finder is provided in Appendix~\ref{sec:fof_example}.

In general, an identified cluster includes young star particles, as well as both inert and active RsvP.
The existing star cluster mass $M_\text{ex}$ is simply the summation of all star particles, while the total reservoir mass $M_\text{rsv}$ includes both the inert and active RsvPs 
in the cluster because both RsvPs will begin to spawn stars before the next FoF step. Another quantity we can obtain from the cluster finding is an array of the mass of existing young stars in this cluster $\{m_{\text{ex}}\}$. In principle, we can broadcast these quantities to all the RsvPs in each group so that they can access their current environmental information for the subsequent IMF sampling. 

For purposes we will explain in the next section, we additionally perform a density estimation among the RsvP after each cluster finding and record the particle IDs of the densest RsvP in each group. 

\subsection{Semi-deterministic sampling: a combined method}
\label{sec:sdt}
Once we obtain $M_\text{ex}$, $M_\text{rsv}$, and $\{m_{\text{ex}}\}$, we can assign the masses of the stars formed in the near future from the $M_\text{rsv}$ reservoir. Assuming the existing stars already follow an IMF defined by $M_\text{ex}$, we can determine the mass of new stars by following these steps:
\begin{enumerate}
    \item Firstly, we calculate the maximum member star mass $m_\text{max}$ and the normalization factor $k$ of the new IMF defined by the total mass $M_\text{tot}=M_\text{ex}+M_\text{rsv}$ using equations~(\ref{eq:mmax}) and (\ref{eq:norm}).
    \item Then, we solve $\{m_i\}$ and  $\{m_{i,\text{low}}\}$ consecutively following the optimal sampling \citep{2017A&A...607A.126Y:Yan}, where $\{m_i\}$ is the center of weight of the IMF bin $i$, and $\{m_{i,\text{low}}\}$ is the lower boundary of the IMF bin $i$.
    \item Finally, for a given mass interval $[m_{i,\text{low}},m_{i-1,\text{low}})$, we check if there is already an existing star in this bin. If not, we need to form a new star with $m_i$ from the RsvPs. Otherwise, we just skip this mass bin. By doing this, we can obtain an array $\{m_\text{new}\}$, which contains a series of unique masses and $\Sigma m_{\text{new},i} = M_\text{rsv}$. 
\end{enumerate} 

This procedure would ensure that the sampled stellar population recovers the assumed IMF, even if different stellar groups (with different $m_{\rm max}$ values) merge at the time step when their distance becomes smaller than the linking length.

In principle, the mass of every single star that spawns from the RsvPs can be calculated deterministically in this way. However, this reasoning encounters two problems in numerical practice. Firstly, it is nontrivial to assign the sampled mass to each RsvP. Determining which sampled mass belongs to which RsvPs requires additional model assumptions.
%\yan{["nontrivial" is not a problem. What is the problem? E.g. My problem is how do we know which sampled mass belongs to which reservoir particle. Do we need mass conservation for each particle or only for the entire star cluster?]} 
Secondly, recording and broadcasting the ${m_\text{ex}}$ and ${m_\text{new}}$ are highly demanding on memory and communication.

A possible way to solve the first is to use an algorithm like {\sc subfind} \citep{2001MNRAS.328..726S:Springel} to further identify the substructures of the RsvP groups and assign the mass at the center of self-gravitating cores accordingly. However, this will introduce additional assumptions and computational costs.
\begin{figure*}
	\includegraphics[width=2\columnwidth]{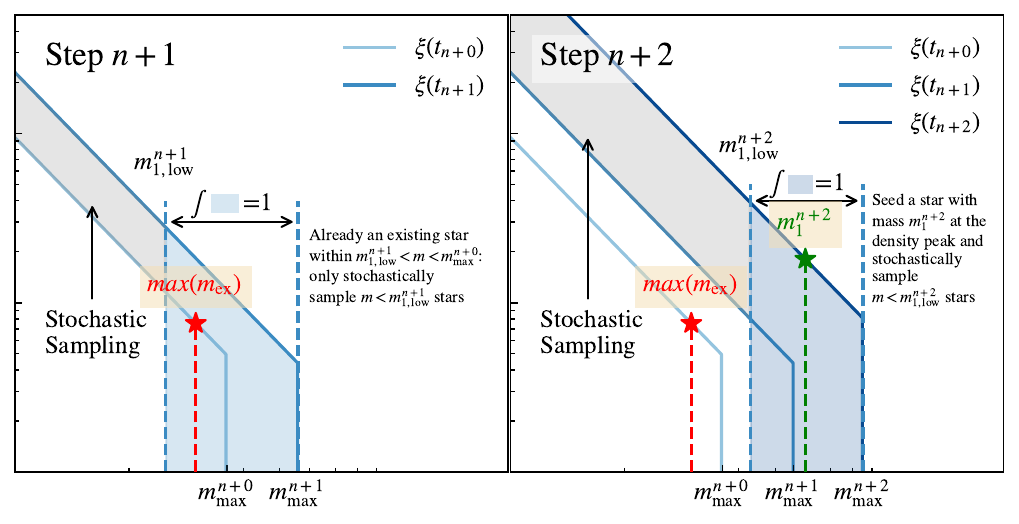}
    \caption{Schematic illustration of the semi-deterministic (SDT) stellar mass sampling method. At each FoF time step $n$, the most-massive allowed star $m_1^n$ is determined from the IMF $\xi(m)$ such that exactly one star occupies the mass interval $[m_{1,\mathrm{low}}^n,\, m_{\max}^n]$, where $m_{\max}^n$ is set by the instantaneous cluster mass $M_{\mathrm{ecl}}^n$. Left panel (Step $n+1$): If an existing star already lies within the updated high-mass interval $m_{1,\mathrm{low}}^{n+1} < m < m_{\max}^{n+1}$, no new massive seed is formed. Instead, only stochastic sampling is performed below $m_{1,\mathrm{low}}^{n+1}$, drawing stellar masses from the inverse cumulative IMF with an upper mass limit of $m_{1,\mathrm{low}}^{n+1}$. 
    Right panel (Step $n+2$):
    If no star occupies the high-mass interval 
    $m_{1,\mathrm{low}}^{n+2} < m < m_{\max}^{n+2}$, 
    a new most-massive star of mass $m_1^{n+2}$ is seeded at the identified 
    density peak of the RsvP group. The remaining mass reservoir is then 
    populated through stochastic sampling below $m_{1,\mathrm{low}}^{n+2}$.}
    \label{fig:sample}
\end{figure*}

Alternatively, we propose an SDT method, which significantly reduces the cost but still satisfies the optimal sampling theory in the observable measures (see Section~\ref{sec:MmR-sim}). 

In Fig~\ref{fig:sample}, we present a sketch of our SDT method to sample stellar mass.  Instead of deterministically assigning the mass and position of all new stars, we only do this for the most-massive star allowed to form at each FoF step. The mass of this star is denoted as $m_1$ and calculated as \begin{equation}
    \begin{cases}
    1 =  &\int_{m_{1,\text{low}}}^{m_\text{max}^n}\xi(m){\rm d}m\,,\\
    m_1^n=&\int_{m_{1,\text{low}}^n}^{m_\text{max}^n}m\xi(m){\rm d}m\,,\\
    M_\text{ecl}^n =& \int_{m_\text{low}}^{m_\text{max}^n}m\xi(m){\rm d}m,
    \end{cases}
\end{equation}
where $m_\text{max}^n$ is the stellar upper mass limit associated with a cluster's instantaneous total stellar mass, $M_\text{ecl}^n$, at a given time step $n$, $m_{1,\text{low}}^n$ is the corresponding lower boundary of the mass bin of $m_1^n$, and $m_1^n$ is the integrated stellar mass within this mass range occupied by the single most-massive star to be sampled at the current time step. As the left panel of Fig~\ref{fig:sample} shown, if at the next time step ${n+1}$ there is already an existing star within the $m_{1,\text{low}}^{n+1}<m_1^{n}<m_\text{max}^{n+1}$ range, we will skip the seed and no longer form a star with mass $m_1^{n+1}$. Instead, we only perform the stochastic sampling following the method outlined by \cite{2024A&A...691A.231D:Deng}, but with an upper mass limit of $m_{1,\text{low}}^{n+1}$. That is, we draw random numbers and map them to the stellar masses by the inverse function of the cumulative function of the IMF, $m_\star=F^{-1}(y)$, where $y$ is a random variable with the value of $[0,1]$ and $F^{-1}$ is the inverse function of
\begin{align}
    F(x)\equiv \int_{m_\text{low}}^{x}\xi(m){\rm d}m/\left[\int^{m_{1,\text{low}}}_{m_\text{low}}\xi(m)dm\right]\ .
\end{align}

In the other case, as the right panel of Fig~\ref{fig:sample} shows, if at time $t_{n+2}$ there is no existing star within $m_{1,\text{low}}^{n+2}<m<m_\text{max}^{n+2}$, we seed a new most massive star from RsvPs at the density peak we found. For the rest of RsvP, we again perform stochastic sampling with an upper mass limit of $m_{1,\text{low}}^{n+2}$. 

After doing this, each active RsvP is given a seed with the mass of the star that will be formed.
If the stellar mass is smaller than its parent RsvP mass, we simply spawn this star from the RsvP and sample a new mass. We sample low-mass stars iteratively until $\Delta t/\Delta t_\text{FoF}\cdot M_\text{rsv}$ stars are spawned or a star more massive than the residual mass of the reservoir is sampled. If the stellar mass is larger than its parent RsvP mass, we transfer the mass from the neighboring RsvPs using the method described in Section~\ref{sec:ngb}.

As in the NGB method, it can occur that there are too few neighboring RsvPs to form the sampled stars. This provides an additional constraint on the location of massive star formation. However, since we want the most-massive star to be formed deterministically, it cannot be omitted due to an insufficient mass of nearby RsvPs. To ensure that the most-massive star is indeed created, we form this star first, prior to sampling all the other stellar candidates. Additionally, we gradually enlarge the search radius $R_\text{search}$ until we locate enough RsvPs. By construction, $R_\text{search}$ always remains smaller than the overall cluster size, since the target mass of the most-massive star is determined by the total cluster mass. This guaranteed search procedure is applied only to the most-massive star, whereas the other stellar candidates that undergo mass transfer are still assigned a fixed value of $R_\text{search}$.

In summary, the SDT employs an on-the-fly FoF algorithm to infer the instantaneous total mass of each cluster (including both existing young stars and nearby RsvP particles) and then generates new stars by deterministically assigning the mass of the most-massive star and randomly sampling the masses of all remaining stars. The free parameters of our model and their fiducial values are listed in Table~\ref{tab:parameters}.

\begin{deluxetable*}{ccccc}
\tabletypesize{\scriptsize}
\tablewidth{0pt} 
\tablecaption{Free parameters in the star formation model \label{tab:parameters}}
\tablehead{\colhead{Function} & \colhead{Symbol} & \colhead{Meaning}& \colhead{Fiducial Value} & \colhead{Description}}
\startdata
{Convert gas to Rsvps}&$n_\text{th}$ & Threshold density & $3\times10^{-21}\,\text{g cm}^{-3}(m_\text{gas}/\Msun)^{-2}$ & Section~\ref{sec:rsvp-form}\\
{}&$T_\text{th}$ & Threshold temperature & $100$\,K & Section~\ref{sec:rsvp-form}\\
{}&$f_\text{J}$ & Jeans factor & $4$ & Section~\ref{sec:rsvp-form}\\
{}&$f_\alpha$ & Self-gravitating factor & $0.5$ & Section~\ref{sec:rsvp-form}\\
\hline
{Convert Rsvps to stars}&$f_\text{inert}$ & Inert time factor scaled to $t_\text{ff}$ & $1$ & Section~\ref{sec:rsvp-form}\\
{}&$R_\text{search}$ & Searching radius for Rsvp accretion & $3$\,pc & Section~\ref{sec:ngb}\\
% {}&$f_\text{min}$ & minimum conversion factor & $0.1$ & Section~\ref{sec:sdt}\\
% {}&$\eta_\text{jet}$ & Jet mass-loading factor & $0.3$\\
\hline
{Determine group properties}&$\Delta t_\text{FoF}$ & Cadence of on-the-fly FoF & $0.1$\,Myr & Section~\ref{sec:cl-find}\\
{}&$L_\text{FoF}$ & Linking length for on-the-fly FoF & $10$\,pc  & Section~\ref{sec:cl-find} \\
{}&$\tau_\text{young}$ & Age cut for young stars & $5$\,Myr  & Section~\ref{sec:cl-find}\\
\enddata
\end{deluxetable*}

\section{Tests in simulations of isolated clouds}
\label{sec:gmcs}
We test the performance of our new SDT  method in simulations of isolated spherical clouds and compare it with the results obtained by the existing methods applying full stochastic IMF sampling (RND) or the mass-conserving NGB sampling. 
The initial conditions (ICs) of clouds are generated using {\sc MakeCloud} \citep{Grudic2021MakeCloud,2022MNRAS.510.4767L:Lane}. The clouds are initially uniform and supported by a Gaussian random velocity field with a power spectrum of $P_v\propto k^{-2}$. The velocity is normalized to achieve a turbulence-supported state with an initial virial parameter $\alpha_\text{turb}=5R_0{\cal M}c_\text{s}^2/(3GM_0)=2$, where $R_0$ and $M_0$ are the initial cloud radius and mass, respectively.

The simulation suite contains seven clouds with different initial masses from $10^3\,\Msun$ to $8\times10^4\,\Msun$ but a similar freefall time of $\sim4$\,Myr. We name the ICs by their initial mass and radius. For example, the cloud with $M_0=2\times10^4\,\Msun$, $R_0=10$\,pc is named as ``M2R10''. For each IC, we run the simulation 15 times with different random seeds to examine the stochastic scatters of different IMF sampling methods. The mass resolution for all the simulations is $1\,\Msun$ and the initial metallicity is $1\,\Zsun$. We list the parameters of the cloud ICs for the simulations in Table~\ref{tab:sims}.  
\begin{deluxetable}{ccccccc}
\tabletypesize{\scriptsize}
\tablewidth{0pt} 
\tablecaption{Molecular clouds initial conditions for the simulations \label{tab:sims}}
\tablehead{\colhead{Mass [$\Msun$]} & \colhead{Size [pc]} & \colhead{$\alpha_\text{turb}$} & \colhead{$t_\text{ff}$ [Myr]} & $t_\text{sim}$ & \colhead{$N_\text{run}$} & \colhead{Name}  }
\startdata
$1000$ & $4$ & 2 & $4.19$ & $20$ & $15$ &M01R4\\
$2000$ & $5$ & 2 & $4.14$ & $20$ & $15$ &M02R5\\
$5000$ & $6$ & 2 & $3.45$ & $10$ & $15$ &M05R6\\
$10^4$ & $8$ & 2 & $3.75$ & $10$ & $15$ &M1R8\\
$2\times10^4$ & 2 & $10$ & $3.70$ & $10$ & $15$ &M2R10\\
$4\times10^4$ & 2 & $13$ & $3.88$ & $10$ & $15$ &M4R13\\
$8\times10^4$ & 2 & $16$ & $3.75$ & $10$ & $15$ &M8R16\
\enddata
\end{deluxetable}
\begin{figure*}
	\includegraphics[width=2\columnwidth]{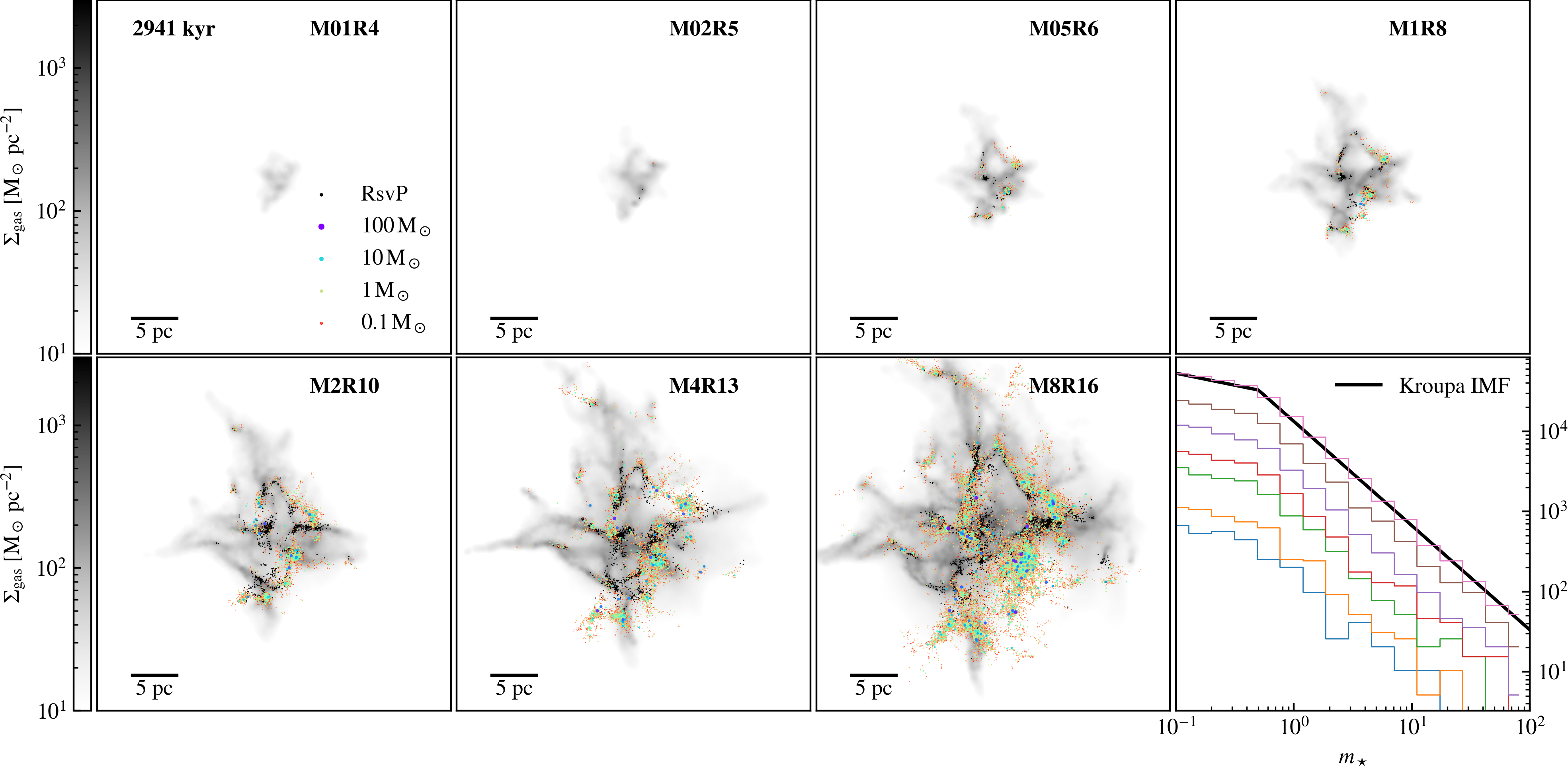}
    \caption{Overview of the simulations of isolated clouds. The first through seventh panels are the snapshots at $3$\,Myr of the seven clouds run with SDT method. The background color maps are the surface density projection, while colorful dots with different sizes are the stars with different masses. The black dots represents the RsvPs. The last panel shows the final IMF of each simulation compared with the canonical Kroupa IMF.}
    \label{fig:sims}
\end{figure*}

To perform the simulations with realistic interstellar medium (ISM) physics, we incorporated the sampling methods to the Realistic ISM modeling in Galaxy Evolution and Lifecycles  \citep[RIGEL][]{2024A&A...691A.231D:Deng} model. The original RIGEL model tracks the evolution and feedback of individual massive stars in the resolved multiphase ISM. We modified the RIGEL model to track the evolution and enrichment of individual low-mass stars in the same way as for massive stars. The coupled system of gravity, magnetohydrodynamics, and radiative transfer is solved using the \arepo code \citep{2010MNRAS.401..791S,2013MNRAS.432..176P:Pakmor,2016MNRAS.455.1134P,2019MNRAS.485..117K,2024arXiv240417630Z:Zier}.

In Fig.~\ref{fig:sims}, we provide an overview of the simulations performed using our SDT method. The snapshots are taken at roughly $0.75t_\text{ff}$. The clouds present filamentary structures as a result of the compression of supersonic turbulence and global contraction under self-gravity. The RsvP particles distribute along the filaments of dense gas. In large clouds such as M4R13 and M8R16, many subclusters have already formed at the intersections of the filaments and knots of dense gas clumps.
Feedback features can also be found in these two clouds. Some gas mass is channeled outward through low-density regions due to stellar feedback, leaving the subcluster exposed. Although we can visually distinguish subclusters within the clouds, these subclusters are usually linked as a single FoF group since we set a rather tolerant linking length of $10$\,pc. The purpose of such a setup is to check whether our functions work without the ambiguity of how to define a cluster. 

As shown in the last panel of Fig.~\ref{fig:sims}, our SDT method recovers a Kroupa-like IMF across the entire stellar mass range. A clear upper mass limit is evident in each simulation and is examined in the following section.

\subsection{Highest stellar mass in clusters}
\label{sec:MmR-sim}

\begin{figure}
	\includegraphics[width=\columnwidth]{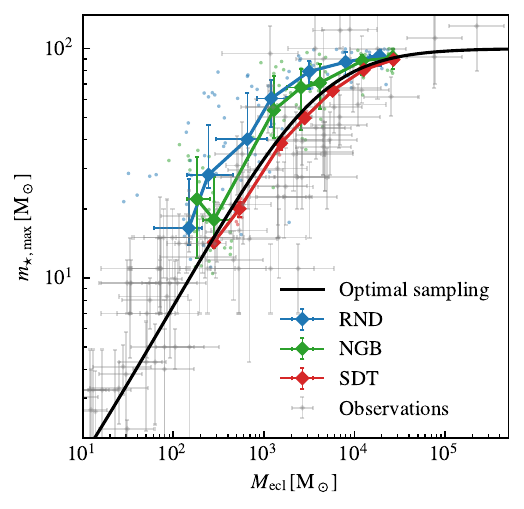}
    \caption{Most-massive stellar mass to embedded cluster mass ($m_{\star,\text{max}}$--$M_\text{ecl}$) relation in the simulations. Blue, green, and red dots are the results of isolated cloud simulations using random sampling (RND), neighbor-based sampling (NGB), and SDT sampling. Gray crosses are the observational data.}
    \label{fig:MmR-sim}
\end{figure}

The $m_{\star,\text{max}}$--$M_\text{ecl}$ relation is the most accessible observational constraint on the correlation between stellar masses and their host cluster mass. In Fig~\ref{fig:MmR-sim}, we compare the performance of the three sampling methods: RND, NGB, and SDT, based on how well they reproduce the observed $m_{\star,\text{max}}$--$M_\text{ecl}$ relation. As the prediction of the optimal sampling theory reproduces the observational statistics, we use it as a reference to compare the models.

Since the original RIGEL model does not explicitly sample stars with masses below $8\,\Msun$, we randomly resample the stellar masses with an imposed upper limit of $8\,\Msun$ for clusters that contain no stars with masses $>8\,\Msun$ in the RND simulations, thereby preventing bias introduced by the mass-cut in the sampling.

By construction, the results using the SDT method closely follow the relation predicted by optimal sampling. The run-to-run variation for a given cloud is negligible, consistent with the observations showing no intrinsic variation in the $m_{\star,\text{max}}$--$M_\text{ecl}$ relation \citep{2023A&A...670A.151Y:Yan}. We emphasize that the SDT method can adjust to a different deterministic sampling rule for the massive stars to reproduce any updated $m_{\star,\text{max}}$--$M_\text{ecl}$ relation from future observational surveys.

On the other hand, both RND and NGB methods show a significantly larger stochastic scatter compared to our SDT method. 70\% of the RND clusters and 52\% of the NGB clusters have the highest stellar mass larger than the prediction of optimal sampling. The stochastic scatter is most prominent for the low-mass clusters. For example, in both RND and NGB models, $30$-$40\,\Msun$ stars appear in clusters with $200$-$300\,\Msun$ masses, while the observed highest stellar masses in similar mass clusters are all lower than $20\,\Msun$ for such clusters.

\begin{figure}
	\includegraphics[width=\columnwidth]{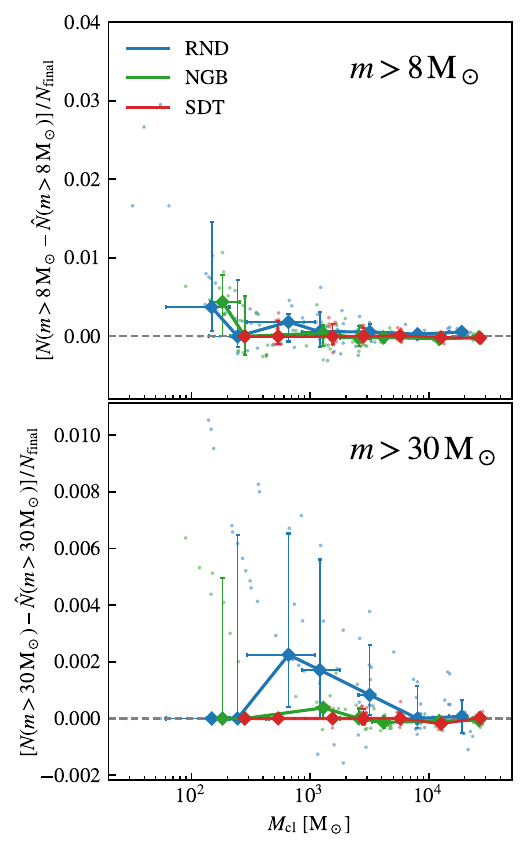}
    \caption{The difference between number of massive stars in the simulation, $N$, and that predicted by the optimal sampling theory, $\hat{N}$, normalized by the total number of stars formed in a simulation, $N_\text{final}$. Each dot is the result of one simulated clouds and the errorbars show the medians and 16-84 percentile ranges. The larger marker points with error bars present the median value and 16 and 84 percentiles for simulated clouds with different IC.}
    \label{fig:Nms}
\end{figure}

In addition to specifying the most-massive star, the optimal sampling theory also predicts the exact number of stars at lower masses. To test whether our SDT reproduces these predictions, we analyze the number of stars exceeding certain mass thresholds. In Fig.~\ref{fig:Nms}, we present the difference between the fractional number of stars more massive than $8\,\Msun$ and $30\,\Msun$ in the simulations and the corresponding values predicted by optimal sampling theory. This metric characterizes the run-to-run variation in the number of massive stars and indicates the extent to which the results deviate from the optimal sampling prediction.

The RND method presents the largest run-to-run variation, and this scatter presents a divergent trend as the cluster mass decreases. For the $m>30\,\Msun$ cut, the scatter becomes zero for the smallest cloud because none of RND simulations of this IC form any star more massive than $30\,\Msun$.

In comparison, the SDT method presents negligible scatter for both $8\,\Msun$ and $30\,\Msun$ mass-cuts across the whole mass range. The median values also closely match the prediction of optimal sampling theory. This result indicates that, even though our SDT model only performs deterministic sampling for the most-massive star at each FoF step, it effectively constrains the sampling process and reproduces the characteristic of optimal sampling in terms of the number of massive stars above different mass limits.

Notably, the NGB method also presents a small run-to-run scatter and closely matches the optimal sampling prediction, even for the $m>30\,\Msun$ cut. This suggests the merit of the NGB method adopted by current state-of-the-art simulations. However, the main run-to-run variation of NGB runs resides in the randomness of the location and timing of massive star and results in scatter in the star formation efficiency (SFE) among runs with the same ICs (see Section~\ref{sec:int_SFE}).

\subsection{Formation time and mass segregation of massive stars}
\label{sec:time-seg}

Including environmental constraints implies that massive stars will generally form only when sufficient reservoirs of star-forming gas are available. Imposing these constraints is expected to produce two outcomes: a delay in the onset of massive star formation and an initial segregation in stellar masses. In this section, we examine these two effects in our simulations.

\begin{figure}
	\includegraphics[width=\columnwidth]{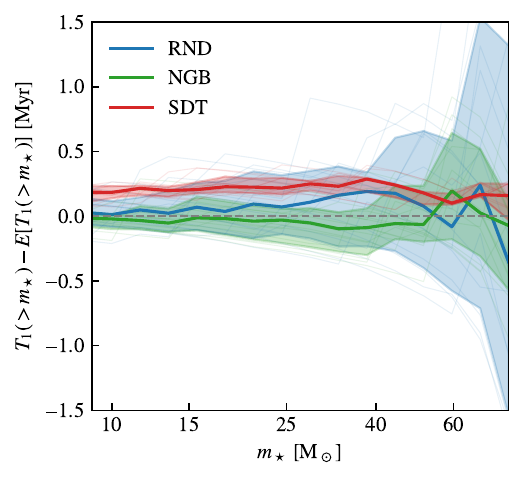}
    \caption{The difference between the formation time of the first star more massive than a given mass threshold ($T_1(>m_\star)$) and its expectation from fully stochastic (Poisson) IMF sampling ($E[T_1(>m_\star)]$) in simulations of the M8R16 cloud. The solid curves and shaded regions show the median and 16th-84th percentile range for the 15 runs with different random seeds. The transparent curves show the results for each individual run.}
    \label{fig:min_tform}
\end{figure}

Recent theoretical studies have suggested that massive stars may form systematically later than their lower-mass counterparts. For example, the global hierarchical collapse model predicts that continued accretion onto dense cores naturally delays the formation of massive stars \citep[][]{2019MNRAS.490.3061V:Vazquez-Semadeni,2024MNRAS.530.3445V}, while radiation-hydrodynamic simulations of clustered star formation likewise find that massive stars assemble their mass over more extended accretion histories \citep[e.g.,][]{2022MNRAS.512..216G:Grudic}. Motivated by these studies, we quantify the delay in massive-star formation predicted by the SDT model and compare it with the expectation from fully stochastic IMF sampling.

In Fig.~\ref{fig:min_tform}, we show the difference between the formation time of the first star more massive than a given mass threshold ($T_1(>m_\star)$) and its expectation from fully stochastic IMF sampling. The expected time is computed by combining the star formation history ($\psi(t)$) with the IMF ($\xi$) to estimate the formation rate of stars above the threshold. The Poisson intensity $\lambda(t)$ is then
\begin{equation}
\lambda(t) = \psi(t)f(>m_\star)= \psi(t)\frac{\int_{m_\star}^{m_\text{up}}\xi{\rm d}m}{\int_{m_\text{low}}^{m_\text{up}}m\xi{\rm d}m}\,.
\end{equation}
For a Poisson process, the probability of having formed at least one star above the threshold by time $t$ is
\begin{equation}
P(\ge 1)=1-\exp\left[-\int_0^t \lambda(t^\prime){\rm d}t^\prime\right]\,.
\end{equation}
We define the expected first-formation time $E[T_1(>m_\star)]$ as the median waiting time, corresponding to a 50\% probability of forming at least one such star. Therefore,
\begin{equation}
\int_0^t \lambda(t^\prime){\rm d}t^\prime=\ln 2\,.
\end{equation}

As an example, we only present the results of the M8R16 cloud. The RND and NGB simulations start forming $>8\,\Msun$ massive stars first, and the median onset time is consistent with a Poisson process. On the contrary, the SDT simulations begin forming massive stars last, with a delay of roughly $0.15$\,Myr compared to the Poisson sampling. For more-massive stars, the time delay tends to decrease, and the medians of different models begin to converge. However, the RND model presents a large run-to-run variation in the formation time of $>40\,\Msun$ stars, and the scatter diverges as the stellar mass increases. To the extreme, the formation time of the first $80\,\Msun$ star can be either $2$\,Myr earlier or $2$\,Myr  later than the expectation, and the 16th--84th percentile interval is as large as $2.8$\,Myr. In contrast, the dispersion in formation times stays low in the SDT runs over the entire mass range, with the 16th--84th percentile interval never exceeding 0.3\,Myr. Such a small scatter means the formation of the first massive star is tightly regulated by the amount of stellar mass formed rather than by stochastic fluctuations. The NGB model presents an agreement with the Poisson expectation and an intermediate scatter with the maximum 16th--84th percentile interval of $0.8$\,Myr.

In addition to the M8R16 cloud used here as an example, the delay in massive-star formation is found in all simulations employing the SDT model, and it shows a negative correlation with the cloud initial mass. In the smallest cloud, M05R4, the median formation time of the first star with $>8\Msun$ occurs 0.5 Myr later than in the Poisson sampling. This stochastic delay may, therefore, contribute to the observed cluster emergence timescale and its inverse dependence on cluster mass reported by \cite{2026NatAs.tmp..112P:Pedrini}; although the observed trend is likely dominated by other physical processes, such as cloud collapse and feedback-driven gas dispersal, which are beyond the scope of the present work.

Another effect of our new IMF sampling method is the emergence of initial mass segregation. Unlike classical mass segregation, which arises from the exchange of momentum and energy among stars in a cluster, initial mass segregation reflects the preferred birth locations of stars. The fact that young clusters are observed with their most-massive stars already concentrated in the center indicates that mass segregation can be an inherent property present from the time of cluster formation \citep[e.g.,][]{2011ApJ...727...64K}. To assess mass segregation in our simulated clusters, we use the mass segregation offset (MSO) as a metric that properly captures the segregation within subclusters as components of the overall hierarchical cluster structure.

Following \cite{2022MNRAS.515..167G}, we define a subcluster as a centrally concentrated stellar overdensity and detect such subclusters using the Variational Bayesian Gaussian Mixture clustering algorithm implemented in the {\sc scikit-learn} library. The MSO is then calculated by
\begin{equation}
    \Lambda_\text{MSO} = \left<\frac{d_\text{subcl}}{R_\text{subcl}}\right>_\text{all}\cdot\left<\frac{d_\text{subcl}}{R_\text{subcl}}\right>^{-1}_\text{massive}\,,
\end{equation}
where $d_\text{subcl}$ is the distance from a star to the center of the nearest subcluster, $R_\text{subcl}$ is the size of the subcluster. The angle brackets denote averaging over all stars or massive stars in the cluster.

\begin{figure}
	\includegraphics[width=\columnwidth]{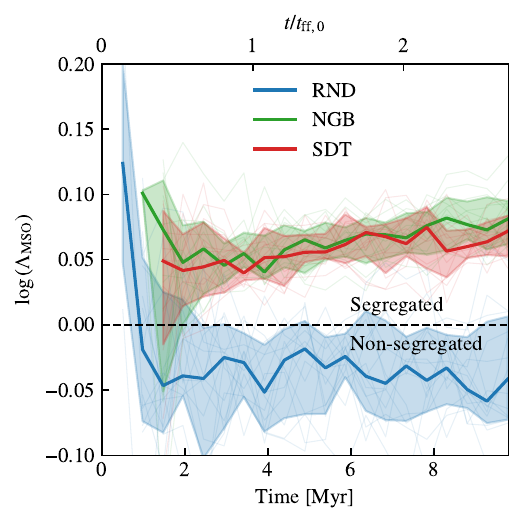}
    \caption{Mass segregation offset of the cluster in simulations of the M8R16 cloud.  The solid curves and shaded regions show the median and 16-84 percentile range for the 15 runs with different random seeds. The transparent curves show the results for each individual runs.}
    \label{fig:segregation}
\end{figure}

In Fig~\ref{fig:segregation}, we exhibit the evolution of MSO in the M8R16 simulations. In the RND simulations, the cluster does not exhibit mass segregation with $\Lambda_\text{MSO}<1$, apart from the initial 1 Myr, since the massive stars are placed at random positions. In contrast, when environmental constraints are introduced, the clusters clearly display mass segregation with $\Lambda_\text{MSO}>1$ in both the NGB and SDT runs. The behavior of MSOs in these two models is very similar, as the SDT model only explicitly specifies the locations of the most-massive stars, while the formation of other massive stars follows the same constraints as the NGB model.

In the NGB and SDT runs, the MSOs show a gradual increase. This can be attributed to dynamical mass segregation, given the several Myr timescale in a $10^3$-$10^4\,\Msun$ cluster \citep{1987gady.book.....B:Binney}. In addition, gas expulsion driven by stellar feedback can further contract the inner regions of subclusters in which the massive stars are already segregated. We also find that the MSO remains constant in the RND runs, which is because all star particles have similar dynamical masses in the original RIGEL model.
\subsection{Feedback luminosity}

\begin{figure}
	\includegraphics[width=\columnwidth]{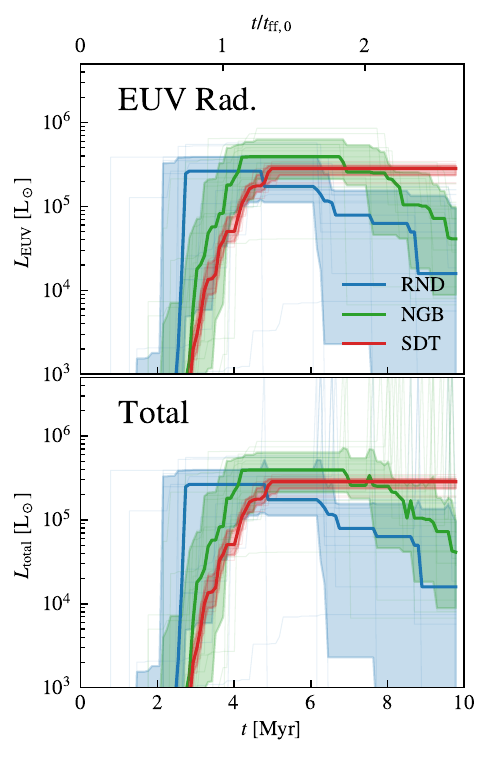}
    \caption{Feedback luminosity as a function of time in the simulated clouds (M05R6). The upper panel plots the evolution of EUV (13.6--100\,eV) radiation luminosity, while the bottom panel plots the evolution of total feedback luminosity as a summation of EUV radiation, stellar wind, and supernovae (SNe). The solid curves and shaded regions show the median and 16th-84th percentile range for the 15 runs with different random seeds. The transparent curves show the results for each individual run. For the SN luminosity, we assume each SN releases $10^{51}$\,erg with 1000\,yr to calculate its luminosity in the dimension of power (energy per time).}
    \label{fig:feedback}
\end{figure}

The number and formation time of massive stars crucially determine the intensity and timing of stellar feedback from the young stars, and the timing of stellar feedback is known to be crucial for regulating both cloud- and galaxy-scale properties \citep[e.g.,][]{2024A&A...681A..28A,2024A&A...691A.231D:Deng}.  Due to the steep mass-luminosity relation ($L\propto M^{3.5}$), the stochasticity in massive star mass significantly affects feedback intensity.  

In the previous section, we showed that the SDT method forms massive stars at later times than the RND and NGG methods, which in turn delays the onset of stellar feedback. To assess the strength of this feedback, Fig.~\ref{fig:feedback} displays how the stellar feedback luminosity evolves in the simulations. We use the results from the $5000\,\Msun$ cloud (M05R6) as an example, since the small number of massive stars can magnify the sampling effects. The stellar feedback is dominated by extreme ultraviolet (EUV; 13.6--100\,eV) radiation in young clusters. Sparse SNe appear in the RND and NGB runs after one freefall time because of the randomly sampled very massive stars, while no SNe occur in the SDT runs over the $10$\,Myr duration of the simulation.

The SDT runs exhibit the smallest variation in feedback intensity from one run to another. This occurs because massive stars emerge in a sequential manner: stars of higher mass are only allowed to form at later times when the cluster has accumulated more mass. While we still apply the stochastic sampling scheme to select most stars, except for the most-massive one at each FoF step, our SDT approach effectively achieves an ordered pattern for the formation of massive stars. In contrast, both the RND and NGB runs show large run-to-run variations.

\subsection{Star formation efficiency}
\label{sec:int_SFE}
\begin{figure}
	\includegraphics[width=\columnwidth]{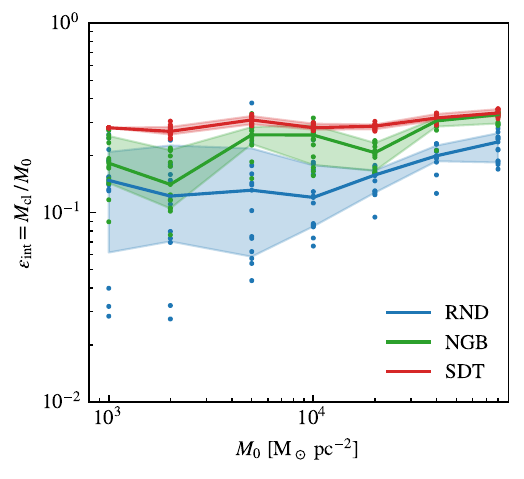}
    \caption{Integrated SFE as a function of cloud initial mass. The solid curves and shaded regions show the median and 16th-84th percentile range for the 15 runs with different random seeds. The dots show the results for each individual run. The SDT runs show the minimized run-to-run variation and highest integrated SFE due to the ordered star formation.}
    \label{fig:intSFE}
\end{figure}
Finally, we examine the integrated SFE ($\epsilon_\text{int}$) of these simulated clouds. The integrated SFE is defined as the ratio of the final mass of stars formed (i.e., the cluster mass $M_\text{cl}$) to the initial gas mass of a cloud, $M_0$.
That is, $\epsilon_\text{int}=M_\text{cl}/M_0$.  

Recent numerical simulations have shown that the integrated SFE is jointly affected by the inward gravitational collapse and the outward stellar feedback \citep[e.g.,][]{2018MNRAS.475.3511G:Grudic,2019MNRAS.487..364L,2025A&A...704A.240D:Deng}.
The more regulated sampling methods avoid the formation of unrealistically massive stars in small star clusters.
Thus, a higher SFE and a lower run-to-run variation are expected in NGB and SDT runs with delayed and ordered massive star formation.

Figure~\ref{fig:intSFE} presents the integrated SFEs of clouds with different masses for the three star sampling methods. As expected, the SDT runs show minimal run-to-run variation and the highest SFEs. The variation among M01R4 runs using SDT methods is nearly zero because all these runs form only two massive stars in a deterministic way. For more-massive clouds with a couple of massive stars, the run-to-run variations are still much smaller than those in the NGB and RND methods. This result indicates that cloud-scale star formation in the SDT model is governed mainly by the hydrodynamic conditions, whereas the choice of random seeds for the stochastic sampling has only a negligible impact on the outcomes.

The run-to-run variations in the NGB and RND runs both show a clear divergent trend as the cluster mass decreases. The SFEs in massive clouds are convergent, as the large number of stars eliminates the stochasticity, whereas in small clouds, the IMF sampling stochasticity begins to become important. 

The median SFE values obtained with the three methods differ by at most a factor of 2.5. The lowest SFE in the RND runs can be an order of magnitude smaller than in the SDT runs. For the NGB runs, the largest deviation from the SDT runs is about a factor of 3.

\section{An example of applications in galaxy simulations}
\label{sec:gal}
Equipped with the new IMF sampling methods, we run a suite of galaxy-scale simulations to test the impact of IMF sampling methods on the global and integrated properties of a galaxy. We select the dwarf--dwarf merger system reported in \cite{2025A&A...704A.240D:Deng} as our test field for models. \cite{2025A&A...704A.240D:Deng} reported that the total SFR of the galaxy changes by a factor of 130 within 200\,Myr between the pre-merger phase and the merging event. The slope of the cluster initial mass function (CIMF) also changes from -2.23 to -1.98. The rapid and drastic change in star formation activities provides a large dynamic range for our tests.
\subsection{Simulation overview}
\label{sec:overview}
\begin{figure}
	\includegraphics[width=\columnwidth]{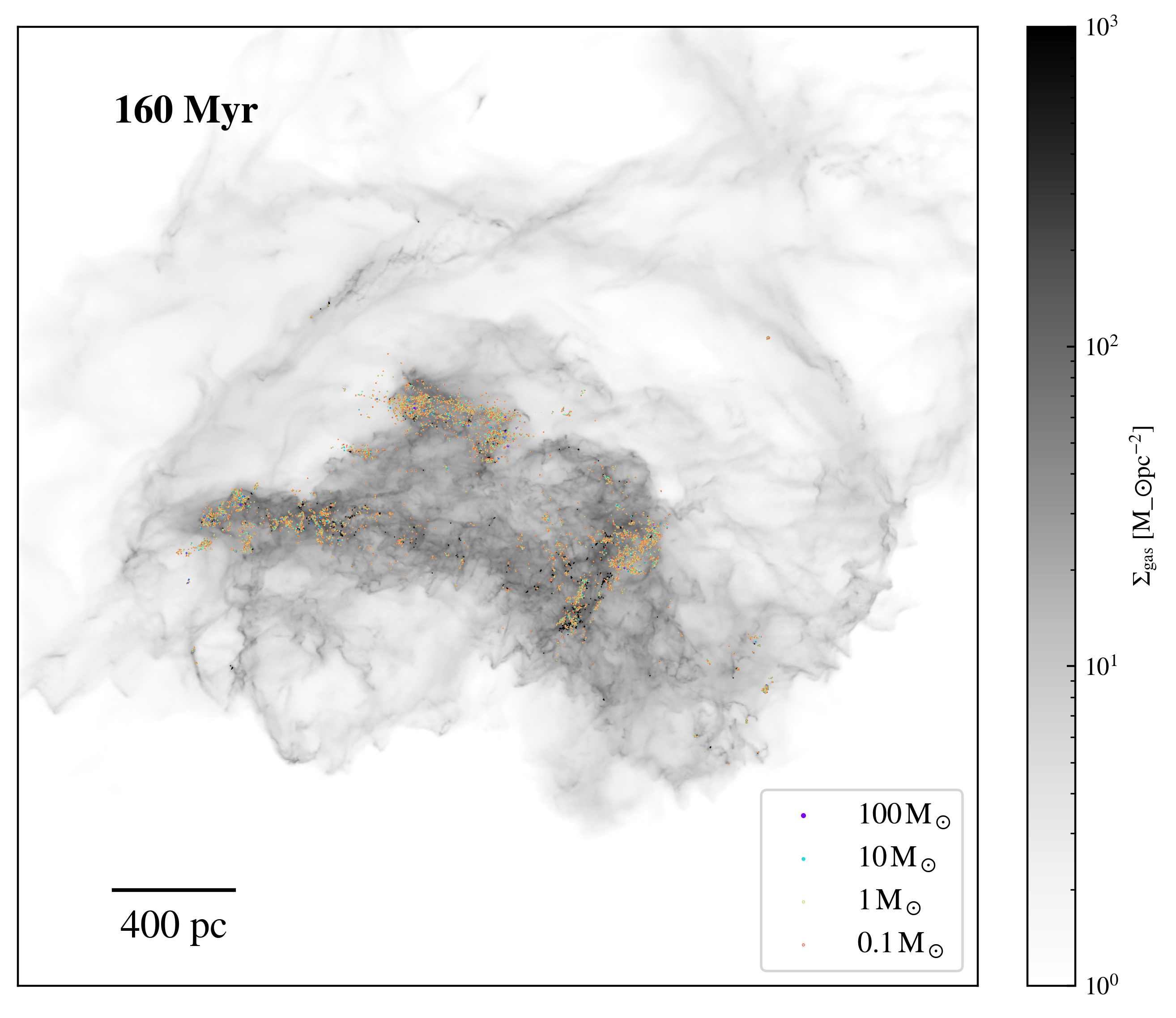}
    \caption{Overview of the merger system at 160\,Myr simulation time. The result is obtained with the SDT model. The background grayscale map is the gas surface density projected along the x-axis, while the colorful dots with different sizes are the young stars (age $<5$\,Myr) with different masses.}
    \label{fig:prj}
\end{figure}
The merger IC is composed of two identical dwarf disk galaxies. Each galaxy contains $2\times10^{10}\,\Msun$ of dark matter (DM), a gaseous disk of $4\times10^7\,\Msun$ with a scale length of $1.46$ kpc, 
and a stellar disk of $2\times10^7\,\Msun$ with a scale length of $0.73$ kpc and a scale height of $0.35$ kpc.
Before assembling the isolated galaxies into the merger IC, we first evolve the isolated galaxy for 200\,Myr with the RND model to make it reach a self-regulated state. The combined merging system is then evolved for an additional 200\,Myr for each sampling method. The two galaxies have two pericentric passages during the 200\,Myr simulation time: the first at $\sim45$\,Myr, and the second at $\sim150$\,Myr. The initial metallicity is set as $0.1\,\Zsun$.
The simulation is performed using the modified RIGEL framework, as introduced in Section~\ref{sec:gmcs}.

We set the mass resolution of gas particles as $10\,\Msun$ to run the simulation with NGB and SDT models. This resolution is lower than the original $2\,\Msun$ setup, but it is still sufficient to achieve converged results on galactic scales \citep{2024A&A...691A.231D:Deng}. We adopt the fiducial star formation parameters listed in Table~\ref{tab:parameters} to run the NGB and SDT simulations. The results for the RND model are obtained directly from the existing simulation snapshots used in \cite{2025A&A...704A.240D:Deng}.

In Fig.\ref{fig:prj}, we show an overview of the simulation using SDT model during the merger. The strong tidal forces drive a substantial quantity of gas into the central area of the merging system, triggering intense star formation. Each individual star converted from star-forming reservoirs is now explicitly sampled and modeled, providing the basis to study star cluster evolution in a galactic environment in the future.

\begin{figure}
	\includegraphics[width=\columnwidth]{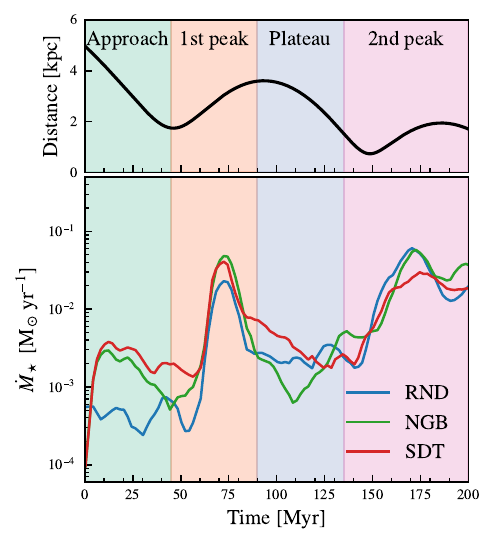}
    \caption{Evolution of the distance between two merging dwarf galaxies (top panel) and the SF history of the entire merger system (bottom panel). The distance is measured between the centers of mass of the background stars in the galaxies, and the SFR is measured as an average over a timescale of 5 Myr. The blue, green, red curves are the results for the RND, NGB, and SDT runs, respectively.}
    \label{fig:SFR}
\end{figure}

Figure~\ref{fig:SFR} shows the evolution of the distance between the centers of mass of disk stars in the two galaxies (top panel) and the global SFR (bottom panel). The dynamics of galaxies in different sampling models have negligible differences.
We, thus, only present the curve of the RND model for illustration. On the contrary, the SFRs show evident deviations, especially when the two galaxies are approaching.

During the approaching stage, the NGB and SDT runs exhibit SFR values that are 4 to 10 times higher than those in the RND run.
This discrepancy is anticipated since the relatively isolated dwarf galaxies tend to host smaller molecular clouds and star clusters, thus reducing the number of massive stars. As discussed in Section~\ref{sec:int_SFE}, the NGB and SDT approaches lead to higher integrated SFE values. The higher SFR here can be regarded as the result of this effect on a galaxy scale. This SFR enhancement in NGB and SDT models is similar to the finding in \cite{2023MNRAS.526.1713S:Steyrleithner}, which truncates the IMF based on cluster mass.

We further notice that the SFRs in the NGB and SDT runs exhibit an overshoot and then decline over time, while the SFR in the RND run remains comparatively steady. This behavior arises because our IC is constructed based on the RND model. Switching the IMF sampling method, however, disrupts this self-regulation by modifying the cloud-scale SFE, which in turn leads to a nonstationary SFR. 

\subsection{Highest stellar mass in clusters within the dwarf merger}
\label{sec:mms-gal}
\begin{figure}
	\includegraphics[width=\columnwidth]{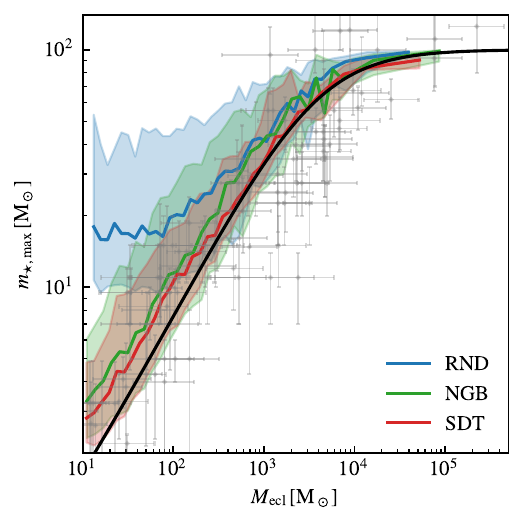}
    \caption{Most-massive stellar mass to embedded cluster mass ($m_{\star,\text{max}}$--$M_\text{ecl}$) relation in the simulations. The clusters are identified by the 4D FoF method. Blue, green, and red curves are the median results for simulations using RND, NGB, and SDT methods, respectively. The shaded regions are the 16th-84th percentile ranges. Gray crosses are the observational data. 
    }
    \label{fig:mms-merger}
\end{figure}
To evaluate the performance of our IMF sampling models, we examine the $m_{\star,\text{max}}$--$M_\text{ecl}$ relation.

Directly applying the standard FoF algorithm simply reproduces the result shown in Fig.~\ref{fig:MmR-sim}, as intended by the SDT method. Therefore, we instead adopt a more physically motivated four-dimensional 4D FoF algorithm, as introduced by \cite{2024A&A...681A..28A} and \cite{2024A&A...691A.231D:Deng}, to identify rapidly evolving clusters in galaxy-scale simulations. It links stars in the ${\bm x}_\text{form}$--$t_\text{form}$ space, where ${\bm x}_\text{form}$ and $t_\text{form}$ are the formation coordinates and formation time of star particles recorded in the simulation snapshots. Stars that are physically connected are linked as a group using a linking length $l_\text{link}$ and a linking time $t_\text{link}$. The resulting group mass should represent the true initial mass of stars that were born in close proximity in both space and time, and it should be unaffected by subsequent dynamical or stellar evolution. We choose $l_\text{link}=5$\,pc and $t_\text{link}=5$\,Myr motivated by the observed size and age spread of star clusters. 

The 4D FoF algorithms perform cluster identification in post-processing based on the simulation snapshots. The cluster identification criteria used in 4D FoF algorithms are stricter than the on-the-fly FoF grouping criteria.
Thus, the results can provide a robustness test of our on-the-fly cluster formation model. In Fig.~\ref{fig:mms-merger}, we show the $m_{\star,\text{max}}$--$M_\text{ecl}$ relation obtained by the 4D FoF algorithms. Similar to what we have done in Section~\ref{sec:MmR-sim}, we resample individual stars from these SSP particles whenever an identified cluster lacks members with $M>8\,\Msun$ in the RND simulation. 

The NGB and SDT models reproduced the observed relation better. For the 4D FoF clusters, the RND method results in a deviation from the optimal sampling curve even for $>10^4$ massive clusters. This deviation is most prominent for clusters less massive than $1000\,\Msun$, and the relation becomes flat within $10\,\Msun<M_\text{ecl}<100\,\Msun$. We note that for $M_\mathrm{ecl}<100\,\Msun$, the mass of the most-massive member star can even exceed the cluster mass. This is because the cluster mass is defined as the sum of the dynamical masses of its member star particles, while the stellar mass shown here corresponds to the feedback mass, which can exceed the dynamical mass of an individual star particle in the RND method because local mass conservation is not enforced. The NGB and SDT models have similar performance, while the SDT result is closer to the optimal sampling curve and presents less scatter. 

We find that the SDT results slightly deviate from the $m_{\star,\text{max}}$--$M_\text{ecl}$ relation at the low-mass end and exhibit a larger scatter than in Fig.~\ref{fig:MmR-sim}. This indicates uncertainties in the identification of clusters when comparing the on-the-fly FoF in 3D position space with the 4D FoF in the ${\bm x}_\text{form}$--$t_\text{form}$ space. In this case, NGB and SDT provide similar performance in simulations with more realistic galactic environments. Thus, the improvement of using SDT sampling is limited for $m_{\star,\text{max}}$--$M_\text{ecl}$ relation. However, we will soon see in the next section that the SDT method is still the only way to reproduce the influence of member star-cluster correlation on the galaxy-wide IMF (gwIMF) predicted by the integrated galactic IMF (IGIMF) theory.

\subsection{Galaxy-wide initial mass function}
\label{sec:gwIMF}
\begin{figure*}
	\includegraphics[width=2\columnwidth]{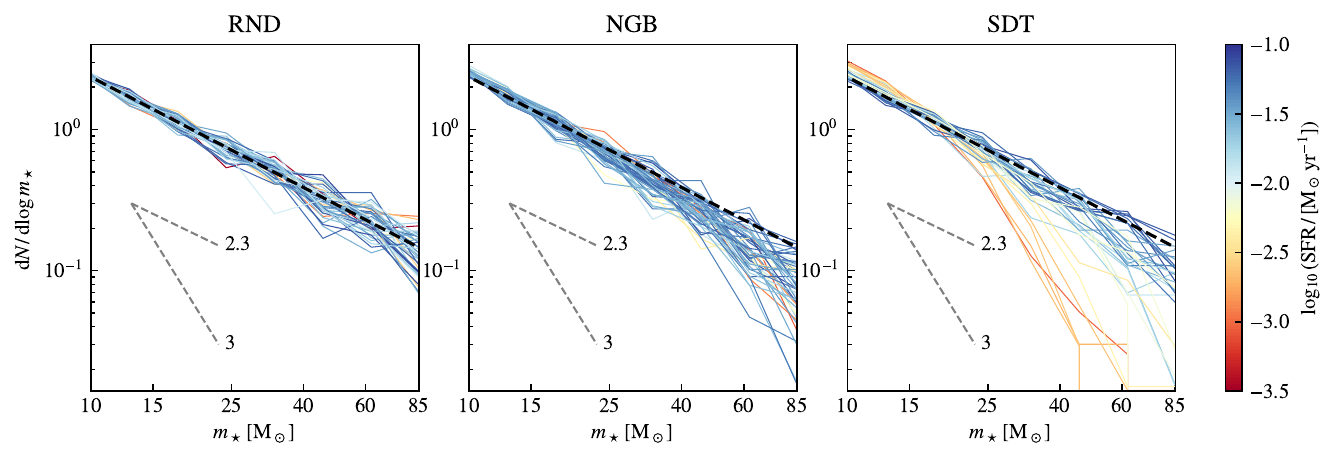}
    \caption{High-mass end of the gwIMFs of the simulated galaxies. The left, middle, and right panels are for the RND, NGB, and SDT models, respectively. The IMFs are color-coded by their corresponding SFR of the galaxy.}
    \label{fig:IMFs}
\end{figure*}
The gwIMF in galaxies with low SFRs was first predicted to be top-light \citep{2003ApJ...598.1076K:Kroupa} and was subsequently confirmed by observations of galaxy photometry \citep{2009ApJ...706..599L} and chemical evolution models \citep{2020A&A...637A..68Y:Yan,2021NatAs...5.1247M}. However, it has been overlooked by the simulation community because of the technical difficulties.

In Fig.~\ref{fig:IMFs}, we show the gwIMFs of the simulated galaxies, color-coded by the average SFR of the galaxies. 
To emphasize the effect, we plot only the high-mass end of the gwIMF. We calculate the IMF for every $5\times10^4\,\Msun$ newly formed star and calculate the corresponding SFR by $5\times10^4[\Msun]/\Delta t$, where $\Delta t$ is the time taken to form these $5\times10^4\,\Msun$ stars. We consider the combined SFR of the two galaxies, since their gas components mix rapidly during the strong interaction. Although this can overestimate the SFR by up to a factor of two, this discrepancy is small compared to the 2 orders of magnitude variation occurring over the course of the merger. 

In the RND model, the gwIMF shows negligible dependence on the SFR. In contrast, the SDT model exhibits a clear variation with the SFR. For low star formation rates, the gwIMF in the $m_\star\gtrsim15\,\Msun$ regime becomes significantly steeper,
exhibiting a slope of $-3$ instead of the canonical value of $-2.3$. The truncation of gwIMF also shifts from $100\,\Msun$ to $\lesssim50\,\Msun$. As the SFR rises, the gwIMF slope becomes progressively flatter and approaches the canonical value. The gwIMF variation can be observed in the NGB model as well, while it has little correlation with the SFR.

\begin{figure}
	\includegraphics[width=\columnwidth]{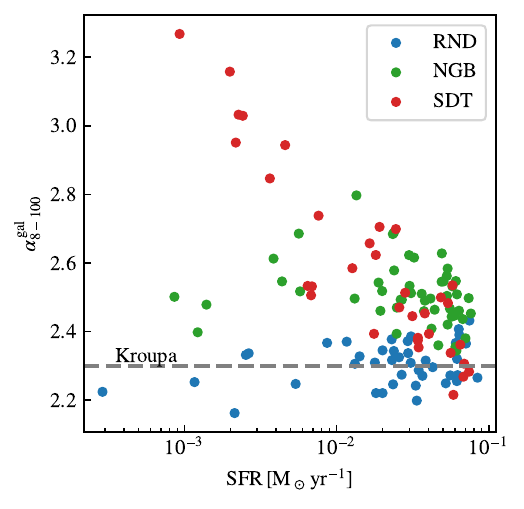}
    \caption{High-mass end gwIMF slopes $\alpha_{8-100}^\text{gal}$ as a function of their corresponding SFR. The slope presents a strong negative correlation with SFR in the simulation with the SDT model.}
    \label{fig:alphas}
\end{figure}
In Fig.~\ref{fig:alphas}, we exhibit the high-mass end gwIMF slopes $\alpha_{8-100}^\text{gal}$ as a function of their corresponding SFR. The power-law index $\alpha_{8-100}^\text{gal}$ is fitted for the $8$--$100\,\Msun$ range using a maximum likelihood estimation. A clear negative correlation between the gwIMF slope and SFR can be seen in the SDT model. At a low SFR of $\sim10^{-3}\,\Msun\,\text{yr}^{-1}$, the high-mass end gwIMF slope even reaches $-3.27$. Except for two data points that roughly lie on the $-2.3$ slope, most of the gwIMF slopes are steeper than the canonical \cite{2001MNRAS.322..231K} value. In contrast, the gwIMF slope only varies between $-2.2$ and $-2.4$ in the RND model. The gwIMF in the NGB model is steeper than the \cite{2001MNRAS.322..231K} IMF as well, while it does not show a strong correlation with SFR and reaches $<-3$ in the steepest case. 

In summary, these results show that only the SDT model can self-consistently reproduce a top-light gwIMF in low-SFR galaxies. The implementation of self-consistent gwIMF variation is of great importance in simulating feedback and chemical enrichment in the dwarf galaxies where a top-light gwIMF is suggested by observations \citep[e.g.,][]{2013ApJ...778..149M:McWilliam,2015MNRAS.446.4220R:Romano,2018MNRAS.477.5554W:Watts,2020MNRAS.495.3276L:Lacchin,2024A&A...681A..76R:Rautio,2026arXiv260610558S:Salvador}.

\subsection{H$\alpha$ as a star formation indicator}
Lastly, we present an example of the impact of IMF sampling methods on the key properties of galaxies. The nebular recombination line H$\alpha$ is commonly employed as a tracer of SFR both in the nearby and high-redshift Universe. The SFR of a galaxy can be derived from the observed H$\alpha$ luminosity via an H$\alpha$-to-SFR conversion factor $C$ calibrated by observations and stellar population synthesis models \citep[e.g][]{1998ARA&A..36..189K,2017PASA...34...58E}, i.e. $\text{SFR} = C\cdot L_{{\rm H}\alpha}$.

\begin{figure}
	\includegraphics[width=\columnwidth]{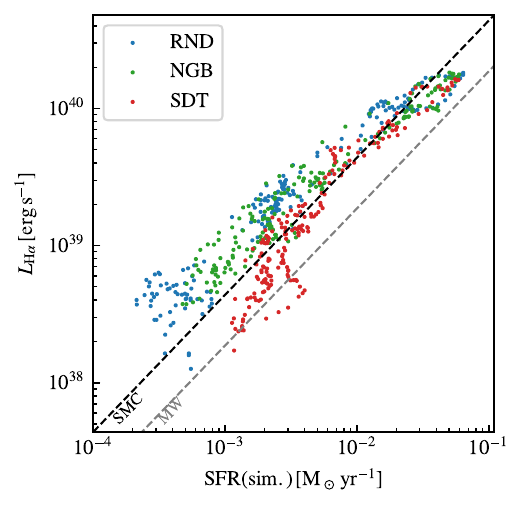}
    \caption{H$\alpha$ luminosity surface density as a function of total SFR of the system. The blue, green, and red dots are for the RND, NGB, and SDT models, respectively. The diagonal lines are plotted to illustrate the Large Magellanic Cloud (LMC) and MW conversion factors.
    }
    \label{fig:Halpha}
\end{figure}

However, as we have seen in the last section, the high-mass-end gwIMF slope can change significantly as a function of SFR, a constant H$\alpha$-to-SFR conversion factor can underestimate the SFR in low-SFR episodes \citep[][]{2007ApJ...671.1550P:Pflamm-Altenburg,2009MNRAS.395..394P:Pflamm-Altenburg,2018A&A...620A..39J:Jerabkova}. To evaluate this effect, we derive the nebular H$\alpha$ emission following the method outlined in \cite{2022MNRAS.514.4560L}. The H$\alpha$ recombination and collisional excitation of hydrogen in warm, nondiffuse gas (i.e. $T<10^{5.5}$\,K, $n>10$\,cm$^{-3}$) are calculated by equations~(3) and (4) from \cite{2013ApJ...779....8K}. To get the intrinsic SFR, we consider young stars with ages 2--10\,Myr. The lower age cut mimics the typical timescale of the embedded phase of young clusters \citep[e.g.,][]{2021ApJ...911..128K,2024ApJ...974L..24D:Deshmukh}, while the upper limit mimics the typical stellar age traced by H$\alpha$.

The resultant H$\alpha$-SFR relation is presented in Fig.~\ref{fig:Halpha}. As a comparison, we plot the H$\alpha$-to-SFR conversions for Small Magellanic Cloud
 (SMC) metallicity ($0.1\,\Zsun$) with ${\log_{10}}C = -41.64$ from \cite{2017PASA...34...58E} and the classic factor for MW-like galaxies of ${\log_{10}}C = -41.27$ from \cite{2012ARA&A..50..531K}. 

At SFRs higher than $0.01\,\Msun$\,yr$^{-1}$\,kpc$^{-2}$, all three models roughly match the relations predicted by the SMC conversion factor. However, at lower SFRs, the relation in the SDT model undergoes a marked downward turn. The conversion factor increased to the MW-like galaxy value, meaning that the H{\sc ii} regions and young star populations are dimmer than those in the other models. The overall distribution differs from the SMC relation by a factor of about $2$-$3$. This result suggests that the H$\alpha$--SFR conversion factor, assuming an invariant IMF, can underestimate the intrinsic SFR by a factor of $2$--$3$ in this case. To the extreme, at fixed $L_{{\rm H}\alpha}$ the intrinsic SFR in different sampling models can deviate by more than a factor of $40$ at maximum, and we expect this effect to be more prominent in lower-SFR environments.

In contrast, the result of the NGB model presents a similar distribution to that of the RND model. These two models do not exhibit any downturn behavior at low SFR. On the contrary, the specific H$\alpha$ luminosity even tends to increase as the SFR decreases. 
This feature is possible due to the variation in the intrinsic properties of the ISM. The SFR variation is driven by the large-scale galaxy merger in our simulations. Thus, the low-SFR episode corresponds to an approaching stage when the ISM is more diffuse, while the high-SFR episode corresponds to the merging stages when the ISM is denser.

We note that, in addition to galaxy-wide IMF variations, stochastic IMF sampling can naturally reduce the ionizing photon output in low-mass stellar populations \citep[e.g.,][]{2011ApJ...741L..26F}, while bursty star formation histories can produce similar signatures because H$\alpha$ and FUV luminosities trace different stellar age ranges \citep[e.g.,][]{2009ApJ...706..599L,2012ApJ...744...44W:Weisz}. These scenarios are not necessarily mutually exclusive, and all may contribute to the observed scatter and systematic trends.
As a detailed calibration of the H$\alpha$-to-SFR conversion factor is beyond the topic of this paper, we simply emphasize that both the observations and numerical simulations suggest that the correlation between stars and their parental clusters is an ingredient that should be taken into account when interpreting galaxy-wide observables and inferring the intrinsic physical properties of galaxies.

\section{Discussions and conclusions}
\label{sec:diss}
\subsection{Implications for galaxy formation models}
In the previous sections, we explored the effect of IMF sampling methods on the properties of simulated molecular clouds and galaxies. The results have several implications for current galaxy formation models and their future development.

In recent years, an increasing number of models have been able to account for individual stars within entire galaxies. These models are powerful tools for studying the cosmological formation of low-mass galaxies \citep[e.g.][]{2022ApJ...941..120G,2025ApJ...978..129A:Andersson}. Although intense starbursts can be triggered at times, these low-mass galaxies could lack the ability to gather sufficient gas locally to assemble massive stars for most of their lifespan. With a lower specific number of massive stars and a lower specific feedback luminosity, star formation activity could be more moderate rather than bursty, as found in previous simulations \citep{2019MNRAS.490.4447W:Wheeler}. Consequently, more diffuse dwarf galaxies with cuspy DM halos could be formed in a moderate star formation mode, thus reconciling the current small-scale challenges of galaxy formation. We emphasize that the halo growth and star formation history of dwarf galaxies can be complicated across cosmic time, and the cosmological simulation with our new IMF sampling method is necessary to actually test these scenarios.

Another population application of such high-resolution simulations is the study of star cluster evolution in galactic environments \citep{2020ApJ...891....2L,2022MNRAS.509.5938H,2025MNRAS.538.2129L:Lahen}. For these simulations, the exact mass spectrum of stars, along with their formation time and coordinates, is of great importance for the evolution of clusters. The delayed feedback from massive stars also postpones the expulsion of gas in the SDT model, causing it to occur when the cluster has reached a higher mass. The delayed gas expulsion may leave a longer time window for the cluster to evolve before the gravitational potential is altered by gas removal, potentially allowing core collapse to occur first. The new IMF sampling models also provide a density preference for the birth position of massive stars. 

Moreover, although our model is built for simulations with solar-mass resolution, only a few galaxy formation models that employ SSP particles at lower resolutions have taken the $m_{\star,\text{max}}$--$M_\text{ecl}$ relation into account. In moderate resolution simulations, the SSP particles with mass $10^3$--$10^4\,\Msun$ represent star clusters \citep[e.g.][]{2018MNRAS.480..800H,2019MNRAS.489.4233M}; The $m_{\star,\text{max}}$–$M_\text{ecl}$ relation determines the truncation mass of the IMF associated with each SSP particle when computing their feedback and yields, thereby affecting the evolution of galaxies. For example, \cite{2014MNRAS.437.3980P:Ploeckinger} showed that adopting a truncated IMF for SSP particles in simulated tidal dwarf galaxies leads to a stellar mass that is 6 times higher. In simulations with even lower resolutions, the SSP particles with mass $\gtrsim10^5\,\Msun$ represent a collection of star clusters and field stars in a kiloparsec-scale region. In this case, the IMF slope of each star particle should change as well.

\subsection{Limitations of this work}
The models described in this work are solely focused on the IMF sampling procedure. However, to realistically model the formation and evolution of star clusters, more physical ingredients are required. 

First of all, we treat RsvPs as a black box of dense cores and protostellar disks, assuming an SFE of 100\% at the scale of RsvP. However, the protostellar disks actually provide feedback through disk winds and jets. The materials in the unresolved cores and disks can, thus, actually return to the ambient gas. More importantly, \cite{2021MNRAS.502.3646G:Guszejnov} showed that jet feedback from protostars effectively suppresses the accretion of massive stars and reduces global SFE, especially for small clouds. In RND models, the feedback from stochastically forming massive stars can override the protostellar feedback in terms of regulating cloud-scale SFE. However, protostellar feedback can play an important role in the NGB and SDT models during the early phase of cluster formation, when no massive stars are present. In Section~\ref{sec:int_SFE}, we have seen that in the NGB and SDT models, the small clouds will have a higher SFE than those in the RND models because of the late appearance of massive stars. When protostellar feedback is incorporated, the final SFE may end up lower than the current results. While it is still likely to remain higher than in the RND model because feedback from massive stars is significantly stronger, a reliable cloud-scale SFE will require future implementation and calibration of a detailed subgrid model for protostellar feedback.

Moreover, we still treat the stars in a collisionless fashion. However, the collisional dynamics can affect the feedback and star formation activities on both cloud and galaxy scales. For instance, massive runaway stars ejected from their birth clusters can leave the dense, star-forming gas before they explode as SNe, thereby reducing the disruption of their parent clouds while enhancing the large-scale outflows from the galaxy \citep{2017ARA&A..55...59N,2020MNRAS.494.3328A:Andersson}.
A simple velocity-kick prescription, such as assuming a power-law distribution for the initial velocities \citep{2011MNRAS.414.3501E,2016A&A...590A.107O:Oh,2023MNRAS.521.2196A}, can be used to model this effect. However, once this kind of effective model is applied, ensuring conservation of momentum and energy becomes nontrivial. An alternative way is to couple the collisional $N$-body dynamics into a hydrodynamics code. However, obtaining an exact $N$-body solution demands $O(N^2)$ computational complexity and extremely small integration time steps, whereas galaxy-scale simulations typically involve at least $10^8$ particles and exhibit complex, peculiar evolution. Very recently, several models have attempted to reconcile this issue using different numerical methods \citep{Dinnbier&Walch2020FLASH,Fujii+2021aSIRIUS,Polak+2024aTorch,Jo+2024EnzoN,Lahen+2025aSPHGAL+KETJU}. Our upcoming {\sc Arepo-CD} framework (Kong et al. in prep.) incorporates a novel integration framework and a high-accuracy few-body integrator based on the slow-down algorithmic regularization (SDAR) method \citep{Wang+2020SDAR}. Equipped with {\sc Arepo-CD}, we will finally enable the self-consistent modeling of the dynamical evolution of clusters in the RIGEL framework.

Lastly, we use an IMF formalism with fixed slopes (equation~\ref{eq:xi}) to study the  sampling effects. Therefore, by construction, we can only obtain a top-light IMF in galaxies with low SFRs. Recent studies propose the dependence of the slopes on metallicity and gas clump density \cite[e.g.,][]{2021A&A...655A..19Y,2026RAA....26b5003G}. These variations are necessary to reproduce the observed top-heavy IMF in metal-poor \citep[e.g.][]{2012MNRAS.422.2246M} and high-SFR environments \citep[e.g.][]{2018Natur.558..260Z:Zhang,2024ApJ...970..136G:Guo,2024MNRAS.534..523C:Cameron}.

\section{Summary}
\label{sec:sum}
In this work, we introduce the SDT scheme to sample individual stars in galaxy simulations with solar-mass resolution. This method effectively captures the key observational properties of stellar masses and their parent clusters, $m_{\star,\text{max}}$--$M_\text{ecl}$ relation, based on optimal sampling theory. The core concept of SDT sampling is to deterministically assign the masses of the most-massive stars according to observational constraints while sampling the remaining stars randomly to reduce computational expense. The SDT method is realized in the {\sc arepo} MHD code by implementing the star formation RsvPs and on-the-fly FoF cluster finding. 

We evaluate the performance of our SDT method and compare the
results with those obtained using previous methods (RND and NGB) in simulations of isolated molecular clouds and a dwarf galaxy merger using the modified RIGEL galaxy formation framework. In the molecular cloud simulations, the SDT method closely
follows the $m_{\star,\text{max}}$--$M_\text{ecl}$ relation predicted by optimal sampling, while both the RND and NGB methods show a significantly larger stochastic scatter and deviate from optimal sampling for small clusters. Importantly, although our SDT model only forms the most-massive stars deterministically, it always returns the exact number of $m>8\,\Msun$ or $m>30\,\Msun$ massive stars in the clusters. On the contrary, the RND and NGB models tend to form both a larger number of massive stars and stars of higher masses in small clusters. Moreover, in both the SDT and NGB models, massive stars are formed at later times, so stellar feedback is also delayed, since these methods require the RsvP to first accumulate before massive stars can form. This environmental constraint of massive star formation also results in an initial mass segregation in the cluster. As a consequence of these effects, the SDT simulations show minimal run-to-run variation and higher integrated SFEs, especially for the small clouds.

In the dwarf--dwarf merger simulation, the SDT method once more yields an $m_{\star,\text{max}}$-$M_\text{ecl}$ relation that most closely reproduces the observed trend. Due to the fewer massive clusters when the SFR is low, the gwIMF becomes steeper than $-3$ since fewer massive stars are formed. Thus, we successfully reproduce a top-light gwIMF in a low-SFR environment as a pure sampling effect. The total SFR becomes higher at the same stage compared with the RND simulation because of the lower specific feedback luminosity. This also suggests that the H$\alpha$ indicator assuming a fixed IMF slope may underestimate the SFR in such relatively quiescent star formation activities.

\begin{acknowledgments}
We thank Volker Springel for giving us access to \arepo. HL is supported by the National Key R\&D Program of China No. 2023YFB3002502, the National Natural Science Foundation of China under No. 12373006 and 12533004, and the China Manned Space Program with grant No. CMS-CSST-2025-A10. YD is supported by the National Natural Science Foundation of China under No. 125B2057. ZY is supported by the Postdoctoral Fellowship Program of CPSF under grant No. GZC20252097 and the National Natural Science Foundation of China under grant No. 12203021.
\end{acknowledgments}

\appendix
\section{Illustrative Examples of the On-the-Fly Cluster-finding Scheme}
\label{sec:fof_example}
\begin{figure*}
	\includegraphics[width=2\columnwidth]{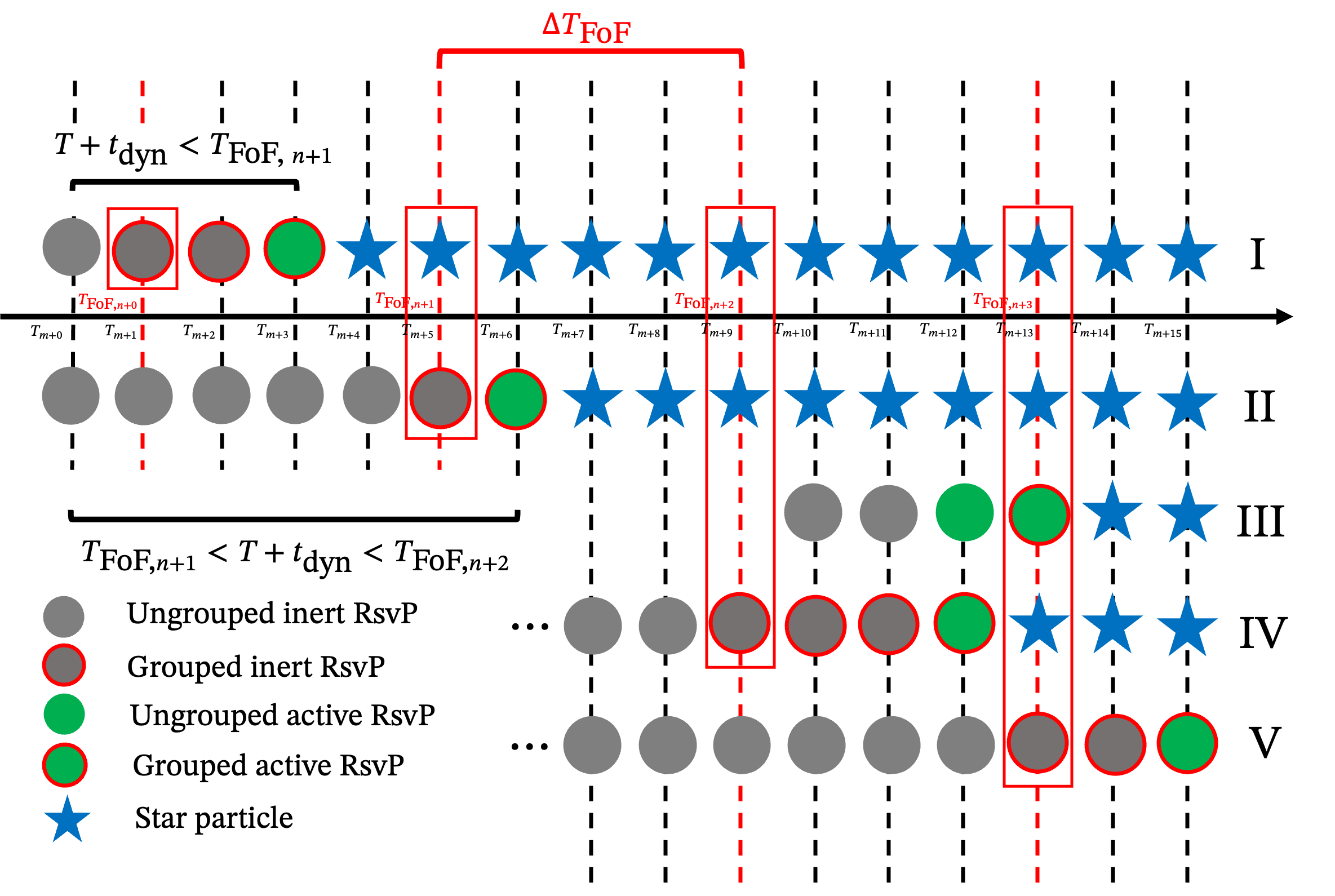}
    \caption{Sketch of the on-the-fly cluster-finding scheme, illustrating different situations in which RsvPs become inert or active relative to the FoF time steps. RsvPs older than their $t_\text{dyn}$ become active to form stars. The five rows represent the evolution of five distinct RsvPs, with each column indicating a particular simulation time step.Only those RsvPs that are activated before the next FoF time step, together with the newly formed stars whose ages are younger than $\tau_\text{young}$, are included in the FoF cluster identification. See section~\ref{sec:cl-find} for details.} 
    \label{fig:fof}
\end{figure*}

Figure~\ref{fig:fof} illustrates the operation of our on-the-fly cluster finding scheme using five representative reserved particles (RsvPs) with different formation times and dynamical times. The five rows demonstrate the possible evolutionary paths of an individual RsvP and how it is incorporated into a star cluster.

Comparing the two RsvPs exhibited in the first (row~I) and second rows  (row~II) of Fig.~\ref{fig:fof}, the one in row~I is expected to become active at the $T_{m+3}<T_{\text{FoF},n+1}$ step. Thus, it will be included in a group at $T_{\text{FoF},n+0}$. On the contrary, the RsvP shown in row~II is expected to become active at the $T_{\text{FoF},n+1}<T_{m+6}<T_{\text{FoF},n+2}$. As a result, it will not be assigned to any group at $T_{\text{FoF},n+0}$ but at $T_{\text{FoF},n+1}$.

In the third row (row~III), an RsvP forms late at $T_{m+9}$. For some reason, it has a very short $t_\text{dyn}$, so it is active before $T_{\text{FoF},n+3}$. In this case, we do not allow it to begin forming stars immediately but keep it inert until it is assigned to a group after $T_{\text{FoF},n+4}$. To avoid the corner case when an RsvP never joins any group, ungrouped active RsvPs will be allowed to form low-mass stars after $\Delta t_\text{FoF}.$

In the fourth row (row IV), an RsvP becomes active right before the next FoF step. In this case, it will be involved in the cluster finding at $T_{\text{FoF},n+2}$. Its mass is accounted for by its parental group, but it is not allowed to spawn stars from itself. Neighboring RsvPs belonging to the same group can borrow mass from this grouped inactive PsvP. If there is still mass remaining, it will begin to spawn stars after it becomes active. Assuming it is entirely converted to stars immediately, then all masses from this RsvP will be involved in the cluster finding at $T_{\text{FoF},n+3}$ as existing stars.

Finally, the fifth row (row V) presents the same situation as row II. An RsvP remains inert for a long time and finally enters the group at $T_{\text{FoF},n+3}$ as it will become active at $T_{m+15}$. The group at $T_{\text{FoF},n+3}$ exemplifies a typical cluster identified by our method, containing young stars as well as both active and inert RsvP.
\bibliography{sdtsample}{}
\bibliographystyle{aasjournalv7}

%% This command is needed to show the entire author+affiliation list when
%% the collaboration and author truncation commands are used.  It has to
%% go at the end of the manuscript.
%\allauthors

%% Include this line if you are using the \added, \replaced, \deleted
%% commands to see a summary list of all changes at the end of the article.
%\listofchanges
\end{CJK*}
\end{document}